\documentclass[preprint,12pt]{elsarticle}

\usepackage{lineno}
\usepackage{epstopdf}

\usepackage[utf8]{inputenc}
\usepackage{amsmath,amsfonts,amsthm} 
\usepackage{physics}
\usepackage{charter}		
\usepackage{amssymb}
\usepackage{graphicx}
\usepackage{bm}
\usepackage{breqn}      
\usepackage{empheq}
\usepackage[toc]{appendix}

\usepackage[breaklinks,hidelinks]{hyperref}
\usepackage{float}        
\usepackage{multirow}
\usepackage{graphicx}
\usepackage[table,xcdraw]{xcolor}
\usepackage{outlines}
\usepackage{svg}
\usepackage{mathtools}  
\usepackage{wasysym}
\usepackage[linesnumbered,ruled,vlined]{algorithm2e}

\usepackage{graphicx}   
\usepackage{subfig}     

\usepackage{comment}        

\newcommand{\qi}[2]{\bm{q}_{#1}^{#2}}             
\newcommand{\qm}[2]{\overline{\bm{q}}_{#1}^{#2}}  

\newcommand{\error}[2]{\bm{\varepsilon}_{#1}^{#2}}             

\newcommand{\xdownarrow}[1]{%
  {\left\downarrow\vbox to #1{}\right.\kern-\nulldelimiterspace}
}

\newcommand{\blocktensor}[1]{\vec{\bm{#1}}}

\begin{document}


\begin{frontmatter}

\title{Accelerating high order discontinuous Galerkin solvers using neural networks: \\ 3D compressible Navier-Stokes equations}

\author[1]{Fernando Manrique de Lara}
\author[1,2]{Esteban Ferrer \corref{cor1}}
\cortext[cor1]{Corresponding author}
\ead{esteban.ferrer@upm.es}

\address[1]{ETSIAE-UPM-School of Aeronautics, Universidad Politécnica de Madrid, Plaza Cardenal Cisneros 3, E-28040 Madrid, Spain}
\address[2]{Center for Computational Simulation, Universidad Politécnica de Madrid, Campus de Montegancedo, Boadilla del Monte, 28660 Madrid, Spain}


\begin{abstract}
We propose to accelerate a high order discontinuous Galerkin solver using neural networks. We include a corrective forcing to a low  polynomial order simulation to enhance its accuracy. The forcing is obtained by training a deep fully connected neural network, using a high  polynomial order simulation but only for a short time frame. 
With this corrective forcing, we can run the low polynomial order simulation faster (with large time steps and low cost per time step) while improving its accuracy.  

We explored this idea for a 1D Burgers' equation in \cite{DELARA2022105274}, and we have extended this work to the 3D Navier-Stokes equations, with and without a Large Eddy Simulation closure model. We test the methodology with the turbulent Taylor Green Vortex case and for various Reynolds numbers (30, 200 and 1600). In addition, the Taylor Green Vortex evolves with time and covers laminar, transitional, and turbulent regimes, as time progresses. 

 The proposed methodology proves to be applicable to a variety of flows and regimes. The results show that the corrective forcing is effective in all Reynolds numbers and time frames (excluding the initial flow development). We can train the corrective forcing with a polynomial order of 8, to increase the accuracy of simulations from a polynomial order 3 to 6, when correcting outside the training time frame. The low order corrected solution is 4 to 5 times faster than a simulation with comparable accuracy (polynomial order 6).  

Additionally, we explore changes in the hyperparameters and use \textit{transfer learning} to speed up the training. We observe that it is not useful to train a corrective forcing using a different flow condition. However, an already trained corrective forcing can be used to initialise a new training (at the correct flow conditions) to obtain an effective forcing with only a few training iterations.   
\end{abstract}


\end{frontmatter}


\tableofcontents

\section{Introduction}
Neural networks (NN) are gaining popularity in scientific computing as they are supplementing/complementing classic fluid mechanics and computational fluid dynamic (CFD) techniques; see reviews by Brunton \textit{et al.} \cite{brunton_kutz_2019}, Garnier \textit{et al.} \cite{GARNIER2021104973} or Vinuesa and Brunton \cite{vinuesa2021potential}, which include a summary of the possible enhancements of fluid dynamic simulations using NN. 

Two trends show promise to exploit the advantages of Neural Networks in the resolution of partial differential equations (PDEs). On the one hand, Physical Informed Neural Networks (PINNS), see the review by Cai \textit{et al.} \cite{Cai2022PhysicsinformedNN}, have been proposed as an alternative to classic Computational Fluid Dynamic methods (CFD). PINNs do not require a mesh and are therefore very flexible and useful for combining a variety of types of data (e.g., experiments and CFD). However, to attain high levels of accuracy, comparable to classic high order methods, it is necessary to include a very wide and deep network, which can limit the usability for turbulent flows. On the other hand, it is possible to combine CFD and NN to accelerate classic methods (e.g., discontinuous Galerkin). Our work belongs to this last category, allowing for fast computations and accurate results.

Combinations of CFD and NN have been proposed mainly for low order methods. Bar-Sinai \textit{et al.}  \cite{BurgersNeuralNetwork}) estimated the spatial derivatives of partial differential equations using NNs for the Burgers' equation on a low resolution grid. In the context of turbulence modelling with NNs, Kochkove \textit{et al.} \cite{Kochkove2101784118} used NN to reconstruct low-fidelity simulation for 2D turbulence, showing the potential of combining CFD and NN. Guastoni \textit{et al.} \cite{Resolution2021} or Güemes \textit{et al.} \cite{GANs} have used NN to spatially interpolate/extrapolate wall quantities in turbulent regimes and generate high-resolution models from low-resolution data. Furthermore, researchers are using NN to derive turbulent closure models or substitute classic LES models, see Stachenfeld \textit{et al.} \cite{Turbulencia} or Beck \& Kurz \cite{Beck} for reviews.

In \cite{DELARA2022105274}, we presented a methodology to accelerate a one dimensional Burgers' equation using NN and a high order discontinuous Galerkin (DG) solver. A corrective forcing, trained with a NN, was included in a low order (low polynomial order) simulation to retrieve high accuracy. 
This corrective forcing decreased the computational cost by reducing the number of degrees of freedom and also by allowing larger time steps (i.e., less stringent Courant–Friedrichs–Levy or CFL restriction). The encouraging accelerations obtained for the 1D Burgers's equations, have lead to this new work where we extend the methodology to 3D Navier-Stokes turbulent flows.  

In high order DG, the physical domain is divided into different elements inside of which the solution is approximated using high polynomial orders. When the polynomial order of the basis is increased, this method shows spectral convergence (for smooth flows), leading to very accurate solutions, e.g., \cite{Kompenhans2016,Kompenhans2016b,Ferrer_2017,Ruedaramirez2019}. An often cited bottleneck of DG methods is that, when increasing the polynomial order, the number of degrees of freedom $dof$ increases to the power of the spatial dimension; typically $dof=N_{el}\times(P+1)^3$ for 3D flows, where $N_{el}$ is the number of elements in the mesh and $P$ is the polynomial order. This rapid increase in $dof$ leads to a non-negligible increase in computational cost. The methodology included in this work bypasses this limitation and allows fast low order simulations that can retain the accuracy of high order simulations.


In this work, we present and apply the methodology to the 3D compressible Navier-Stokes equations in turbulent regimes.
We chose the Taylor Green Vortex case to test the acceleration of the method for different Reynolds numbers. The TGV problem, see Section \ref{sec:TGV} considers the temporal evolution of an initially smooth laminar flow, which transitions to turbulence to follow an isotropic decay. This variety of regimes allows us to test our methodology on smooth, transitional, and fully turbulent flows. Note that in this work we do not consider the effect of walls, since 3D turbulence is already a significant challenge for the methodology. 

When considering turbulent flows, it could be tempting to interpret the change of resolution (or polynomial order), together with the NN correction, as a turbulent subgrid closure model. However, we prefer to think of the method as a correction for low order simulations, without the necessity of defining physical scales and filters. By doing so, we can consider a change of polynomial order in our simulations, which includes not only turbulent effects but also changes of the overall resolution (all scales). This simplification enables the application of the method to all flow regimes including variety of Reynolds numbers and time-evolving flows (with laminar, transitional, and turbulent regimes as time progresses).

The rest of the paper is structured as follows. We present the methodology in Section \ref{SecMethodology}, where we cover the mathematical grounds for the NN acceleration, explain how to obtain the corrective forcing, the selection of time-steps, detail post-processing, and error analysis to finalise with implementation details. %
In Section \ref{SecResults}, we present and discuss the results for the TGV case. We include results for three Reynolds numbers (30, 200 and 1600), for a variety of time frames at $Re=$1600, study the effective acceleration for the method. Finally, in Section \ref{sec:hyperparameters}, we study the effect of hyperparameters and \textit{transfer learning} to obtain an efficient method.
Conclusions and future work are given in Section \ref{SecConclusions}.

\section{Methodology}
\label{SecMethodology}

Let us consider a High Order (HO)  DG discretisations of the 3D NS equations:

\begin{equation}
   \frac{d \boldsymbol {q}_{HO}}{d t} =  \boldsymbol {p}_{HO}(\boldsymbol {q}_{HO}),
    \label{Eq:SystemOfPDEs_HO}
\end{equation}
where $\qi{HO}{} \in  \mathbb{R}^{dof^{HO}}$ contains the state variables of the compressible NS, see \ref{sec:cNS}, including $(  \rho , \rho u ,\rho v , \rho w , \rho e)^T$ for all $dof^{HO}$ in the DG mesh (including the h-mesh and the polynomial p-mesh). The discrete operator $\boldsymbol p_{HO} : \mathbb{R}^{dof^{HO}} \to  \mathbb{R}^{dof^{HO}}$ is a discrete approximation of the NS operator. Here,
$dof^{HO}=N_{el}\times(P^{HO}+1)^3$ denotes the high order degrees of freedom for 3D flows, where $N_{el}$ is the number of elements in the mesh and $P$ is the polynomial order.

We define a filter (or projection) as a lineal operator $\boldsymbol G :\mathbb{R}^{dof^{HO}} \to  \mathbb{R}^{dof^{LO}}$ such that $\qm{HO}{}=\boldsymbol {G} \qi{HO}{}$.  
We apply the filter to Eq. \eqref{Eq:SystemOfPDEs_HO} to obtain an evolution for the filtered high order equation:

\begin{equation}
   \frac{d \qm{HO}{}}{d t} =  \boldsymbol {G}\boldsymbol {p}_{HO}(\qi{HO}{}),
    \label{Eq:SystemOfPDEs_HOfilter}
\end{equation}
where $\qm{HO}{} \in  \mathbb{R}^{dof^{LO}}$ is a low order solution obtained once filtered, and is expected to be more accurate that a low order solution; and  $dof^{LO}=N_{el}\times(P^{LO}+1)^3$. 
Additionally, we define a low order evolution equation

\begin{equation}
   \frac{d \qi{LO}{}}{d t} =  \boldsymbol {p}_{LO}(\qi{LO}{}),
    \label{Eq:SystemOfPDEs_LO}
\end{equation}
where $\qi{LO}{} \in  \mathbb{R}^{dof^{LO}}$ and  $\boldsymbol p_{LO} : \mathbb{R}^{dof^{LO}} \to  \mathbb{R}^{dof^{LO}}$. 

It is easy to see that, in general $\qm{HO}{} \neq \qi{LO}{}$ since they are obtained from different evolution equations. Furthermore, due to the non-linearity of the NS operator $\boldsymbol p$, it is clear that the filtered high order Eq. \eqref{Eq:SystemOfPDEs_HOfilter} and the low order Eq. \eqref{Eq:SystemOfPDEs_LO} do not provide the same solution or accuracy, since 
$\boldsymbol {G}\boldsymbol {p}_{HO}(\qi{HO}{}) \neq \boldsymbol {p}_{LO}(\qi{LO}{})$.

In this work, we propose to find an approximation for the high
order filtered solution $\qi{NN}{} \approx \qm{HO}{}$ using a low order evolution operator which is complemented with a corrective forcing. We rewrite Eq. \eqref{Eq:SystemOfPDEs_HOfilter} with an additional corrective forcing $\boldsymbol{s}( \qi{NN}{}, \qi{HO}{})$ that accounts for the missing scales and interactions between low and high order solutions, and call the new variable $\qi{NN}{}$:

\begin{equation}
   \frac{d \qi{NN}{}}{d t} =\boldsymbol{p}_{LO}(\qi{NN}{})+\boldsymbol{s}( \qi{NN}{}, \qi{HO}{}),
    \label{Eq:SystemOfPDEs_NN}
\end{equation}
where the corrective forcing $\boldsymbol{s}( \qi{NN}{}, \qi{HO}{}) : \mathbb{R}^{dof^{LO}}\times\mathbb{R}^{dof^{HO}} \to  \mathbb{R}^{dof^{LO}}$ is a function of both low and high order solutions and corrects the low order operator to retrieve a high order filtered evolution in time.

Since we seek the same dynamics for $\qi{NN}{} \approx \qm{HO}{}$, we can try to find a corrective forcing that satisfies $  {d \qi{NN}{}}/{d t}={d \qm{HO}{}}/{d t} $. To do so, we subtract Eq. \eqref{Eq:SystemOfPDEs_NN} from Eq. \eqref{Eq:SystemOfPDEs_HO}, and obtain an expression for the corrective forcing:

\begin{equation}
  \boldsymbol{s}( \qi{NN}{}, \qi{HO}{})=
   \boldsymbol {G}\boldsymbol {p}_{HO}(\qi{HO}{})-\boldsymbol{p}_{LO}(\qi{NN}{}),    \label{Eq:SystemOfPDEs_s}
\end{equation}
where $\boldsymbol{s}( \qi{NN}{}, \qi{HO}{})$ is an unknown operator that 
takes into account high order modes and the nonlinear interactions between low- and high-order modes. This term will be modelled through a deep neural network and is key to the proposed methodology.

Eq. \eqref{Eq:SystemOfPDEs_s} is an explicit expression for the corrective forcing that enables the calculation of $\boldsymbol{s}( \qi{NN}{}, \qi{HO}{})$ as long as the high order solution $\qi{HO}{}$ and the low order solution $\qi{NN}{}$ are known. To bypass the limitation of needing the high order solution, we will train a neural network only for a limit amount of time using the high order solution, to then correct the low order solution, after the training is completed. This procedure allows us to extrapolate the corrective forcing and increase the accuracy of a low order  solution, for times where the high order solution is unknown.  



Finally, we consider a temporal disretisation to advance the high and low order corrected systems in time Eq. \eqref{Eq:SystemOfPDEs_HO} and Eq. \eqref{Eq:SystemOfPDEs_NN}. For example,  the corrected system Eq. \eqref{Eq:SystemOfPDEs_NN} becomes:

\begin{equation}
 \qi{NN}{n+1} =\qi{NN}{n}+\Delta t[\boldsymbol{p}_{LO}(\qi{NN}{n})+\boldsymbol{s}^n(\qi{NN}{n}, \qi{HO}{n})],
        \label{Eq:time idscrete}
\end{equation}
where $\Delta t=t_{n+1}-t_{n}$. For simplicity, we have considered explicit Euler time advancement, but the methodology can be easily generalised to other explicit time marching schemes.
In the remaining of this paper, we use a third-order Runge-Kutta scheme.   

There are advantages of using a low polynomial order with a corrective forcing (as opposed to a high polynomial order) that lead to reducing the computational cost: 
\begin{enumerate}
    \item Reduction in degrees of freedom.
    \item In explicit schemes, there is less restriction in the time step due to the CFL (Courant–Friedrichs–Levy condition).
\end{enumerate}
Our methodology takes advantage of these two points by requiring less costly evaluations per time step (fewer degrees of freedom) and fewer number of iterations (less restrictive CFL). 

Note that the decomposition presented in this section is reminiscent of a Large Eddy Simulation approach, where $\boldsymbol{s}$ could denote the subgrid model. However, in our methodology, we prefer to pose the problem in terms of high and low order discretisations, where $\boldsymbol{s}$ can include turbulent effects but also changes in polynomial resolution (not necessarily fine scales). This generality enables us to consider more extreme changes of polynomial orders and to cope with a variety of flow regimes (e.g., laminar, transitional, and turbulent).


\subsection{Neural Network acceleration}
We use a fully connected deep-NN to perform the correlation between the forcing term $ \boldsymbol{s}( \qi{NN}{}, \qi{HO}{})$ and the low order state variable $\qi{NN}{}$. Note that the forcing will also take into account, during training, the effect of the high order solution $\qi{HO}{}$. We define three steps (see Algorithm \ref{Algoritmo}):
\begin{enumerate}
    \item High order evolution (for a short time)
    \item Neural network training (for a short time)
    \item Low order evolution with corrective forcing (for a long time)
\end{enumerate}
\textbf{1. High and low order evolution}. To generate/train the corrective forcing, we need to generate high order data. High order data is expensive to generate (large number of degrees of freedom and small time steps are required) and for this reason we run the high order solver for short time frames. In fact, in this work we generate high order data for long times to be able to evaluate the errors a posteriori (comparing high order filtered solutions with the low order corrected solution), but the high order data are only strictly necessary for short training times. In addition, we generate high order filtered data, which is also needed for training. In summary, during this step, we obtain high order solutions $\qi{HO}{}$ and filtered solutions $\qi{NN}{}=\qm{HO}{}$.


\textbf{2. Training in neural networks}. In this second step, we use a fully connected deep NN to approximate the corrective forcing $ \boldsymbol{s}( \qi{NN}{}, \qi{HO}{})$, using the loss function (explicit expression for $\bm{s}$) defined by Eq.  \eqref{Eq:SystemOfPDEs_s}. The input of the NN is the vector of filtered state variables $\qm{HO}{}$ and the output is the corrective forcing $\bm{s}$ at each time step, which has been trained to fulfill the loss function  Eq.  \eqref{Eq:SystemOfPDEs_s}. The weights obtained during training allow us to evaluate $\bm{s}$ given any low order data $\qi{NN}{}$. More details are provided in Section \ref{sec:force_model}.

\begin{figure}
\centering
\includegraphics[width=\textwidth]{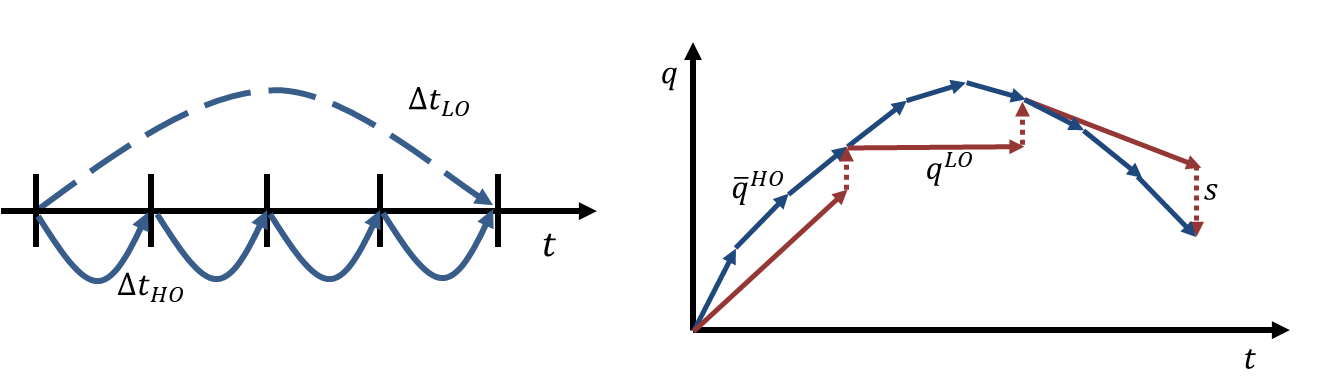}
\caption{Sketch of the time advancement algorithm. Left: The low order time steps (modelled with NN) $\Delta t_n$ are larger than the high order time step $\Delta t_m$. Right: Evolution of the high order variables (blue), which evolve several times to match one single time step from the low order solution (red continuous line). Low order state variables $\qi{}{LO}$ require a corrective forcing $\bm{s}$ to recover high accuracy $\qi{}{HO}$.}
\label{Fig:TimeStepsV2}
\end{figure}

\textbf{3. Low order evolution}. We can now evolve in time the low order simulation using the corrective forcing given by the NN, which only depends on the low order data $\qi{NN}{}$ and was trained to match the forcing term for high order filtered data. Of course, we assume that the corrective forcing will enhance the accuracy of the low order simulation during training, but this would not provide any speed ups. For this reason, we also use the corrective forcing for time evolution outside the training time, which can provide significant speed-ups (see Results section).  Let us remind the reader that evolving the low order solution is much faster since fewer degrees of freedom are needed and a larger time step can be selected (see Figure \ref{Fig:TimeStepsV2} and algorithm \ref{Algoritmo}). 

\begin{algorithm}
\tcp{High order evolution}
\For{n in number of high order iterations}{
match low order with filtered high order $\qi{NN}{n}=\qm{HO}{n}$ \\
save and advance $\qi{NN}{n} \to \qi{NN}{n+1}$ (without correction) \\
\For{m in $\Delta t_{LO} / \Delta t_{HO}$}{
advance $\qm{HO}{m+1}$}
filter $\qi{HO}{n+1}$ to obtain $\qm{HO}{n+1}$ \\
compute and save $\textbf{s}_n = (\qm{HO}{n+1} - \qi{NN}{n+1})/\Delta t_n$}

\tcp{NN training}
Train corrective forcing $\textbf{s}_n$ using a deep NN.

\tcp{Low order evolution}
\For{n in number of low order iterations}{
advance $\qi{NN}{n+1}$ (without correction) \\
compute $\textbf{s}_n$ \\
correct $\qi{NN}{n+1}\leftarrow \qi{NN}{n+1} + \Delta t_n \textbf{s}_n$
}
\caption{Pseudo-algorithm of the proposed methodology. Modified version from \cite{DELARA2022105274}}
\label{Algoritmo}
\end{algorithm}


An additional step (not compulsory) can be added after advancing for arbitrary times the low order solution. This step will recover a high order solution from a low order solution and is not included in the paper since it is not essential to the methodology. We detailed a simplified version of this step in \cite{DELARA2022105274} for the 1D Burgers' equation.  
In 3D more sophisticated reconstructions can be used to recover the high order solution from the low order solution. For example, Generative Adversarial Networks (GANs) have shown promise \cite{GANs}. In what follows we only focus on advancing accurately and fast a low order solution, without taking into account the reconstruction. 


\subsection{Corrective forcing}\label{sec:force_model}
\begin{figure*}
\centering
\includegraphics[width=0.7\textwidth]{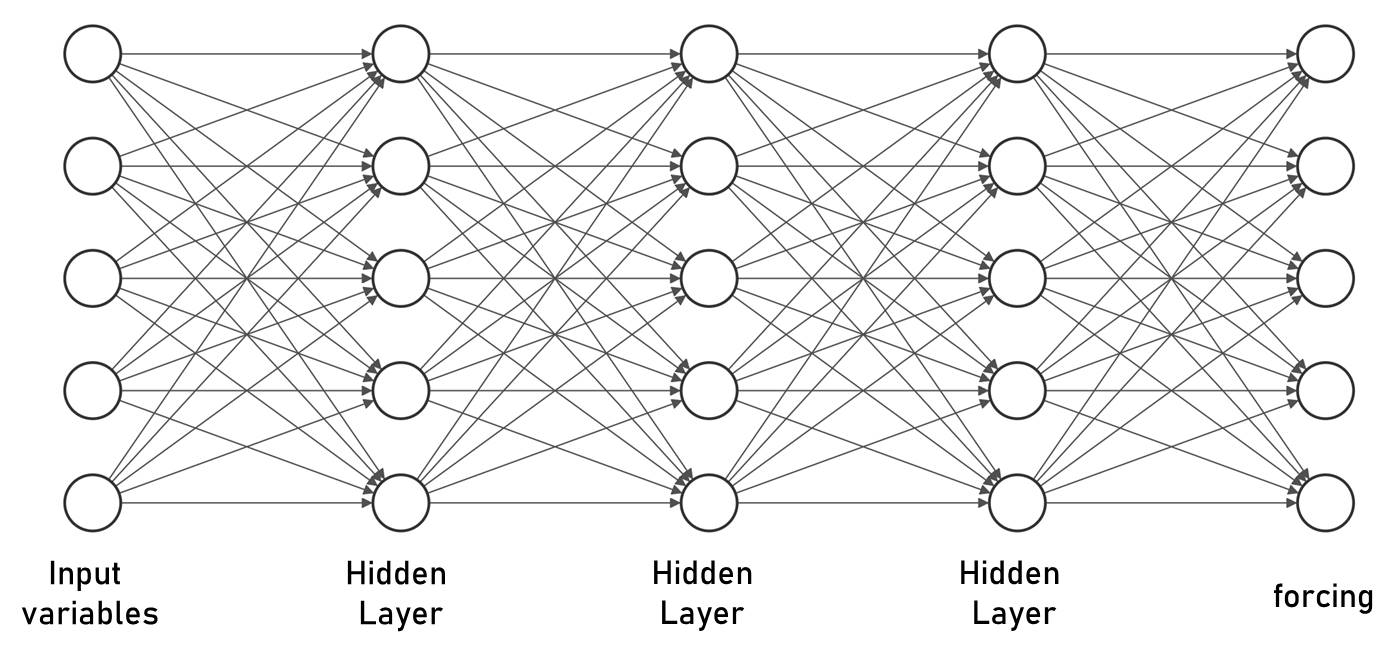}
\caption{Fully connected deep neural network scheme. The input is a vector containing the nodal values for $\rho u$, $\rho v$ and $\rho w$ inside the element, and the output is a vector containing the nodal values for the forcing ($s_2$, $s_3$ and $s_4$) inside the element.}
\label{Fig:RedNN}
\end{figure*}

 We construct a corrective forcing $\bm{s}$ for each degree of freedom in the low order mesh. 
We choose a fully connected Neural Network for its efficiency and simplicity, although other alternatives (convolutional or recursive could have been considered). The NN has $N_{la}=4-5$ layers, and uses for the first and last layers \textit{linear}, while \textit{relu} are used for all intermediate hidden layers, as shown in Figure \ref{Fig:RedNN}.

 We choose to have the same NN architecture for every element and to use all elements to train the network. This is advantageous in the selected turbulent case because the Taylor Green Vortex does not show large variations between elements, since we are dealing with almost homogeneous flow fields (no walls). Therefore, all data can contribute to the training, since an element can represent the flow field of another element shifted in time or space.

Regarding the state vectors, we performed a preliminary test where all variables (density, momentum, and energy) were included in the training but noticed that we could not obtain a valid corrective forcing. Indeed, our test case considers incompressible flows ($Ma=0.08$) leading to a large disparity of scales between density/energy on the one hand and momentum on the other. To avoid this problem, we only use the three components of momentum for the training and re-scale the values in each node. Therefore, the corrective forcing only has three components:
$\bf s=(0,s_2,s_3,s_4,0)^T.$

\subsection{Reduction of degrees of freedom and faster time-steps}
In this section, we estimate the reduction of the number of degrees of freedom and the possibility of using larger time steps. In the remaining of this text, we select $P_{HO}=8$ for the high  polynomial order and $P_{LO}=2$ or $P_{LO}=3$ for the low  polynomial order, depending on the case; see the results Section \ref{SecResults}. 

We can estimate the reduction in the number of degrees of freedom when switching from $P_{HO}=8$ to $P_{LO}=3$, by calculating the ratio between $dof^P={N_{el}\times(P+1)^3}$ for high and low polynomial orders, to obtain $saving_{dof}= \frac{dof^{P=8}}{dof^{P=3}}=(9/4)^3=11.39$. This estimation shows that the potential gain in terms of the number of degrees of freedom is 11. Of course, evaluating the corrective forcing is costly and the effective decrease in cost per time step never achieves the theoretical factor of 11.

Regarding the time step selection, we use explicit time schemes and therefore the time step is limited by the Courant–Friedrichs–Levy number (CFL). In compressible viscous solvers, there are convective and viscous CFL restrictions:

\begin{equation}
    CFL_a \propto \frac{c \Delta t}{\Delta x}(P+1)  \;\;;\;\;
    CFL_v \propto \frac{\Delta t \nu}{\Delta x^2}(P+1)^2.
\end{equation}
Again, we can estimate the potential gains by calculating the ratio between or CFLs for high and low polynomial orders:
\begin{equation}
    \frac{CFL_a^{P=8}}{CFL_a^{P=3}} = 2.25 \frac{\Delta t_{P=8}}{\Delta t_{P=3}}  \;\;;\;\;
    \frac{CFL_v^{P=8}}{CFL_v^{P=3}} = 5.06 \frac{\Delta t_{P=8}}{\Delta t_{LO}}.\label{cfls}
\end{equation}
We can consider the same CFL number (for high and low order polynomial orders) to explore the ratio of time steps and the speed ups associated with using larger times in low order methods. Eq. \eqref{cfls} shows that when switching from P=8 to P=3 we can gain a factor of 2 (advection) and 5 (viscous) in terms of the time step. Note that in practise, the CFL for low order is less restrictive (and can reach one) than for high order and therefore further speed-ups are possible when considering different CFLs for high and low orders. 

\begin{figure*}
 \centering
  \subfloat[\centering $\Delta t_{P}$ for a $8^3$ mesh with $CFL=0.4$.]{
    \includegraphics[width=0.49\textwidth]{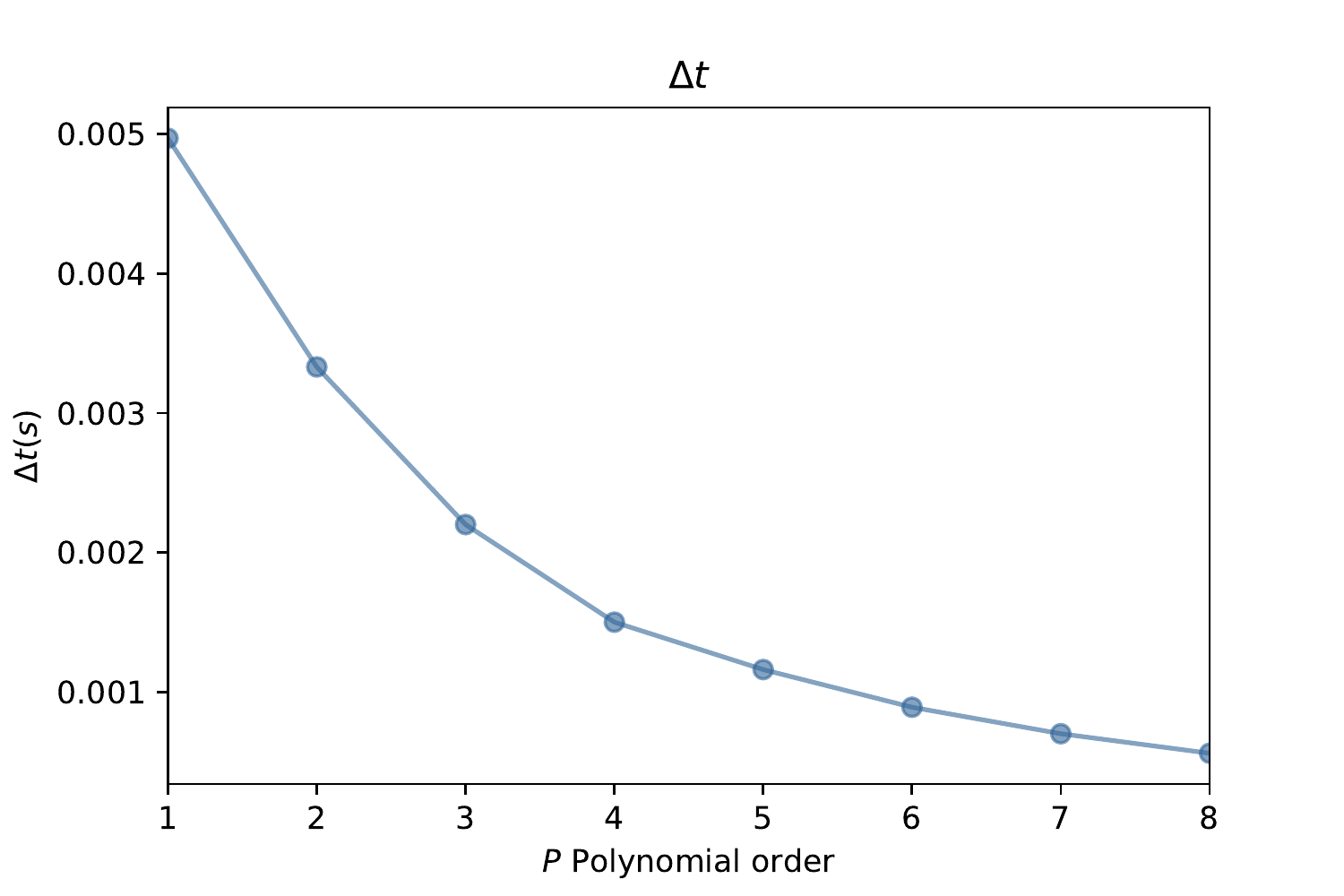}
     \label{Fig:Deltat}}
  \subfloat[\centering Ratio $\Delta t_{P}/\Delta t_{P=8}$ for a $8^3$ mesh with $CFL=0.4$]{
    \includegraphics[width=0.49\textwidth]{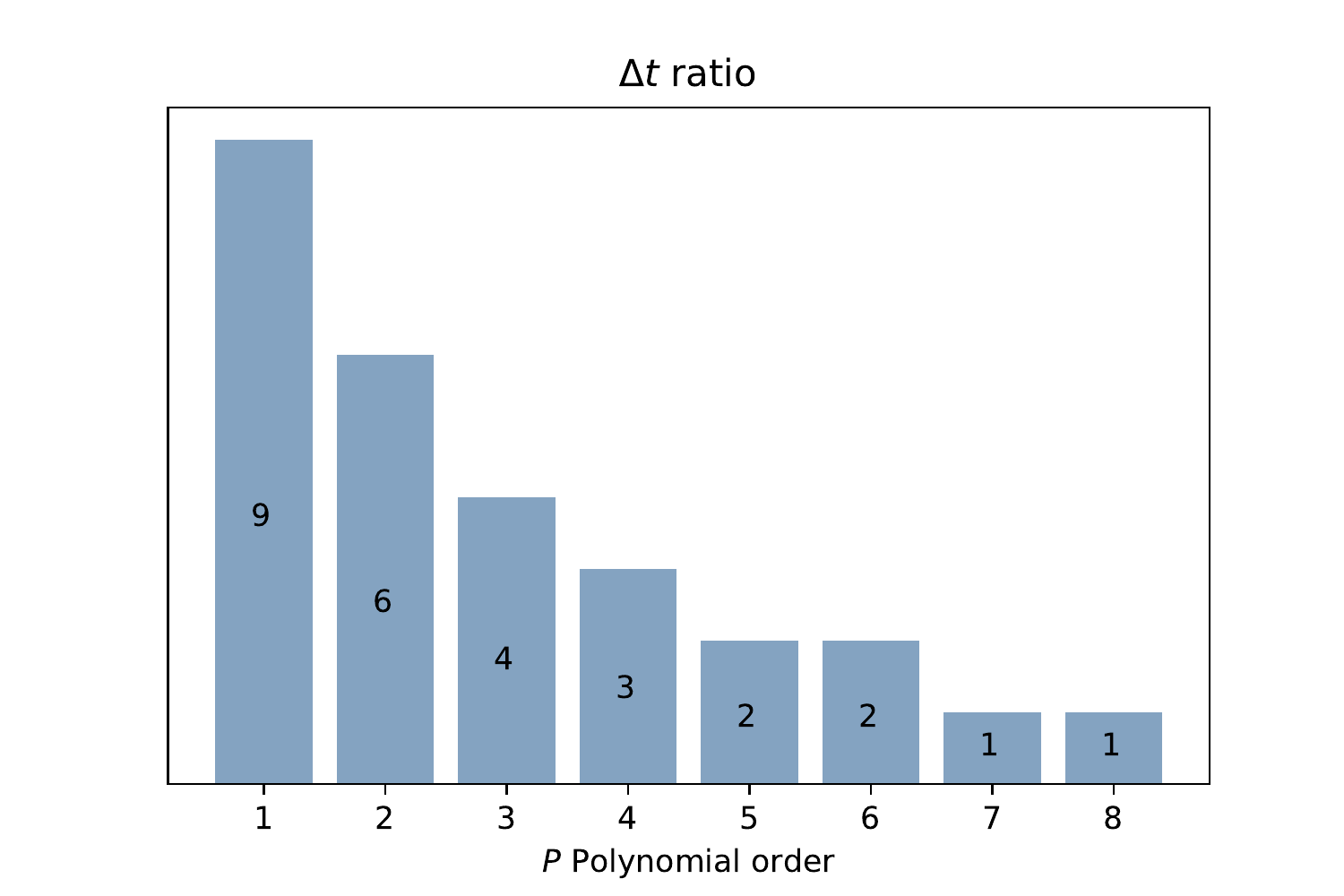}
    \label{Fig:Deltat_ratio}}
    \caption{Temporal time step difference depending on the time step.}
\end{figure*}

Here we fix to $CFL^{P=8}=CFL^{P=3}\approx 0.4$. The next step is to calculate the time step $\Delta t$ for $CFL=4$ for each polynomial order (see Figure \ref{Fig:Deltat}.left) and using Eq. \eqref{cfls} calculate the ratio between high and low order time steps. This is depicted in Figure \ref{Fig:Deltat_ratio}.right), where we truncate the ratio to integer numbers. We can see that if the switch from $P=8$ to $P=1$ we can accelerate by a factor of 9, while when switching from $P=8$ to $P=6$ the ratio is only 2, showing the potential acceleration of our methodology when considering the time advancement speed.

In addition, this procedure allows us to determine how many iterations we must run for the high order simulation to match the time advancement of one low order iteration. Note that the ratio presented is only an estimate since the CFL condition depends on the local flow velocity and should be checked in each stage of the simulation.

\subsection{Post-processing and filtering}

Throughout the results section, we show the time evolution of the infinite norm for the difference between the high order solution $P_{HO}=8$ and low order solutions (with and without corrective forcing) $P_{LO}=2,3$:

\begin{equation}
    \bm{e}_n = \norm{\qi{LO}{n}-\qm{HO}{n}}_{\infty},
\end{equation}
where $n$ denotes the iteration in time. To compute this norm, we filter the high order solution and compare it with the low order solution as time progresses. For each case we include three figures: a) error for the x-momentum $\rho u$, b) error for the y-momentum $\rho v$ and c) error for the z-momentum $\rho w$. We do not present errors in $\rho$ or $\rho e$ because the flow is incompressible; and hence density is constant and the energy equation irrelevant.

\subsection{Error analysis and modelling}\label{sec_error_main}
In \ref{ApendiceError}, we include an error analysis for the NN corrective forcing. 
Based on the error analysis, we propose a simple model to predict the error behaviour as time evolves, when using our methodology. The approximate evolution of the error can be approximated as 

\begin{equation}
    \dv{\norm{\error{}{}}}{t} \approx \alpha \norm{\error{}{}} + \beta,
\end{equation}
where $\error{}{}$ represents the error associated with the difference between the high and low order solution with corrective forcing, $\alpha$ is an effective value that represents the system Jacobian (i.e., the maximum eigenvalue when considering the L2 norm) and $\beta$ is a term independent of the error $\error{}{}$. When including the corrective forcing,  $\alpha$ is associated with the effective Jacobian, which includes two contributions: a Jacobian coming from the NS equations ($\norm{\partial \bm{p}_{LO}/\partial \overline{\bm{q}}}$) and a second Jacobian related to the linearisation of the NN ($\norm{\partial \bm{s}/\partial \overline{\bm{q}}}$), while $\beta$ represents the error of the NN ($\norm{\bm{e}}$) and can be controlled through modifications in the hyperparameters; see \ref{ApendiceError} for details.

\begin{figure*}
\centering
\subfloat[\centering Different values for $\alpha$ and $\beta$]{
\includegraphics[width=0.49\textwidth]{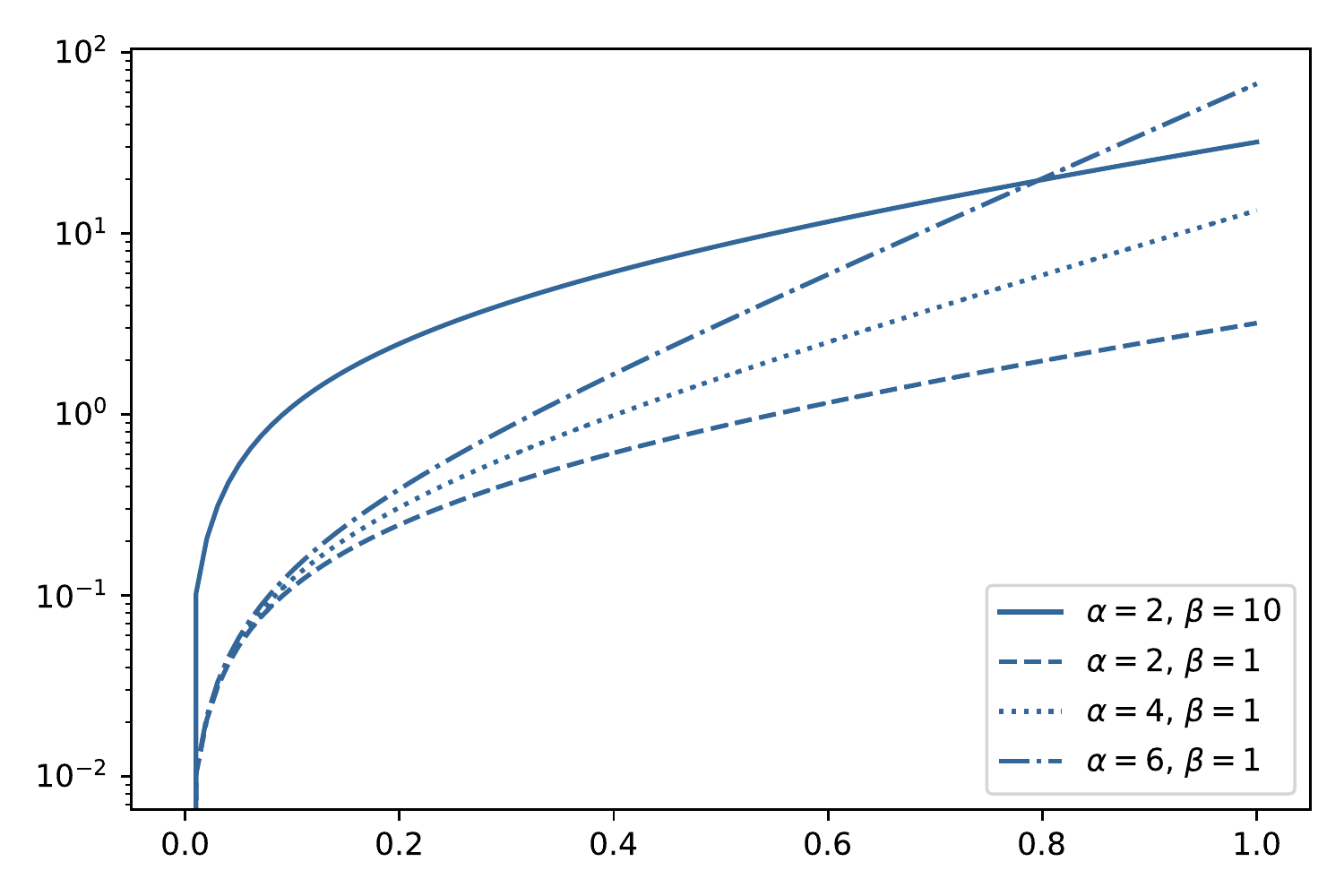}
\label{Fig:CurvesEstimation1}}
\subfloat[\centering Effect of a sudden change in $\beta$ for $t>0.25$]{
\includegraphics[width=0.49\textwidth]{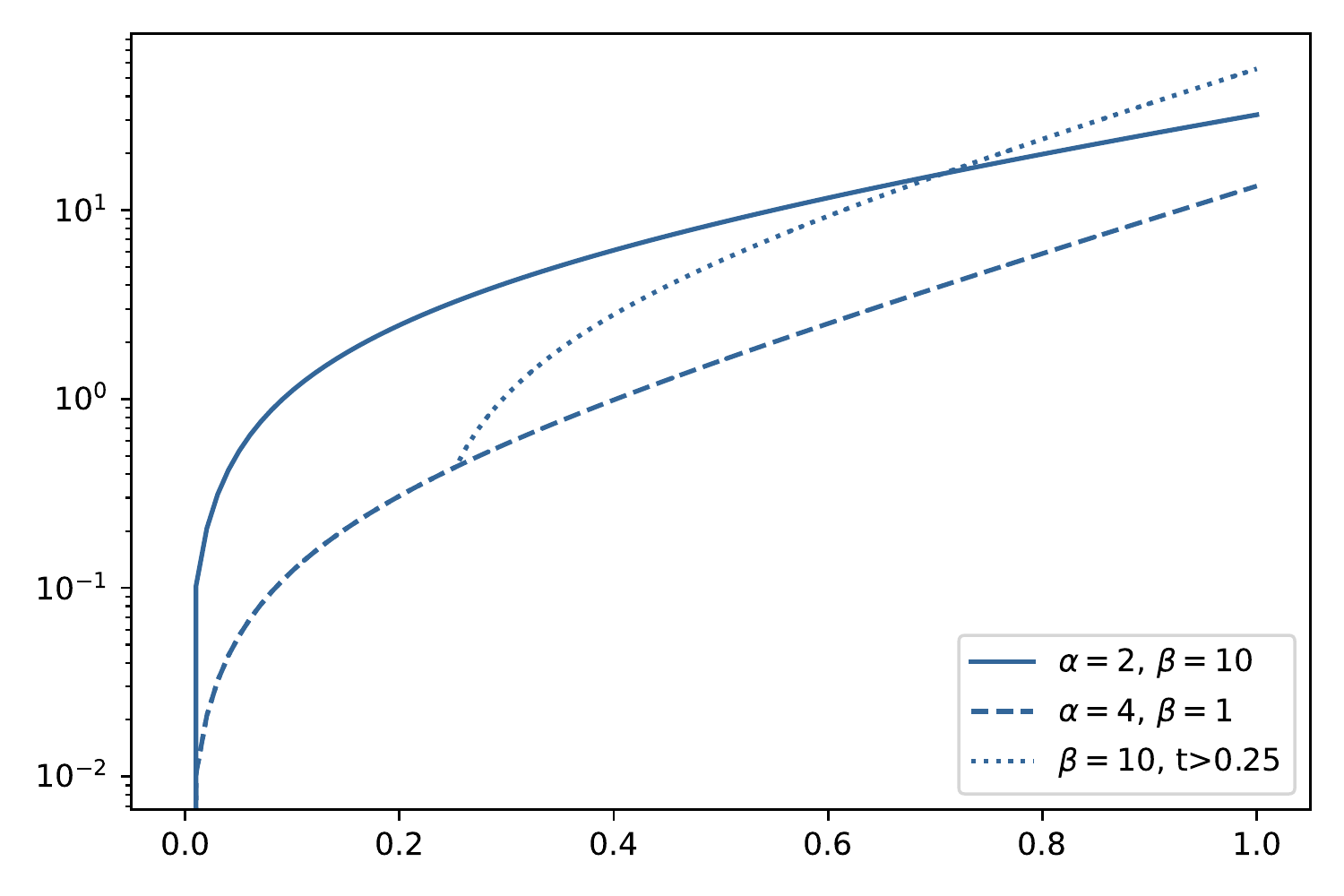}
\label{Fig:CurvesEstimation2}}
\caption{Evolution of the theoretical error over time.}
\end{figure*}

Figures \ref{Fig:CurvesEstimation1} and \ref{Fig:CurvesEstimation2} show error curves for different values of $\alpha$ and $\beta$, and allow us to study the error behaviour as time progresses. 
Figure \ref{Fig:CurvesEstimation1} shows that low values of $\alpha$ lead to a sharper slope in the simulation with corrective forcing, which minimises the usefulness of the corrective forcing, since the error of the corrected scheme soon reaches the noncorrected error. The effect of $\beta$ is to shift the entire curve  down/up. Consequently, we prefer a low value of $\alpha$ and a low value for $\beta$. Note that 
$\alpha$ relates to the system Jacobian (including NS and NN) and hence cannot be modified/controlled easily, while $\beta$ relates to the NN error which can be controlled through modifications in the hyperparameters; see Section \ref{sec:hyperparameters}.

Figure \ref{Fig:CurvesEstimation2} shows the effect of varying $\beta$ for $t>0.25$. This relates to the neural network being a good extrapolating model for the window including the training data, but not afterward. We see that a sudden change in the NN error $\beta$ leads to a sudden increase in errors.

%
%

These simple models can help us to understand the different causes of errors in the figures included in the results section \ref{SecResults}.

\subsection{Implementation Details}
On the one hand, the simulations are run using HORSES3D, a nodal high-order discontinuous Galerkin spectral element method (DGSEM), \cite{2022arXiv220609733F}, written in modern object–oriented Fortran. The background of the DGSEM formulation can be found in \cite{kopriva2009implementing}. One peculiarity of this efficient high order method is the use of nodal Lagrange bases. To perform the filtering necessary in the proposed methodology, we transform from nodal to modal space, filter, and re-project into nodal values. This is a common procedure in DG schemes, and details can be found in our previous work \cite{RUEDARAMIREZ2021109953,MANZANERO2020104440}.

On the other hand, neural networks are evaluated in Python using \textit{KERAS}. To evaluate and import into HORSES3D a trained deep neural networks (weights and biases), we use the library \textit{Fortran to keras bridge} \cite{fortrankeras}. 

While HORSES3D is parallelised with OpenMPI, all Python trainings are run in serial, which is an influencing factor when computing the training times and reported in the result section. All simulations and trainings have been run on a desktop computer with Intel(R) Core(TM) i7 CPU 3.07 GHz and RAM 14,0 GB.

\section{Results and discussions}
\label{SecResults}
\subsection{The Taylor Green Vortex problem}\label{sec:TGV}
Numerical experiments have been performed with the Taylor--Green Vortex (TGV) \cite{TaylorGreen} for a range of Reynolds numbers $Re=$ 30, 200 and 1600, see Figure \ref{Fig:DifferentReynodlsIsoSurface}. The TGV problem has been widely used to study numerical methods to analyse how they can reproduce different flow regimes (for Reynolds numbers large enough$> 200$).
An initial unstable laminar flow, a transitional laminar-turbulent flow, and a fully turbulent flow with isotropic decay, see, for example, \cite{doi:10.1063/1.5083870}. 

The configuration of the TGV problem is a three-dimensional periodic box $[-\pi,\pi]^3$ with the initial condition,

\begin{equation}
\begin{split}
\rho &= \rho_0,\\
v_1 &= V_0 \sin x\cos y \cos z,\\
v_2 &= -V_0\cos x \sin y \cos z,\\
v_3 &= 0,\\
p &= \frac{\rho_0 V_0^2}{\gamma M_0^2} + \frac{\rho_0 V_0^2}{16}(\cos 2x + \cos 2y)(\cos 2z + 2).\\
\end{split}
\label{eq:TGV}
\end{equation}
All simulations are run in a coarse h-mesh with $8^3$ elements and polynomial orders ranging from $P=2$ to $8$, see the mesh in Figure \ref{Fig:BoxMesh}. Details on our particular DG formulations can be found in \cite{2022arXiv220609733F}.  We use Roe \cite{ROE1981357} for advective fluxes and Bassi Rebay 1 \cite{BR1} for viscous fluxes.
We use a third-order three-stage Runge--Kutta scheme to march the equations in time. The flow is incompressible with a Mach number of $M_0=0.08$ in all simulations. 


For Re=1600, we also present results using a Smagorinsky large eddy viscosity model, as this helps us to generalise our methodology to DNS and LES type simulations, with varying polynomials.


%

%
%
%


\subsection{Three Reynolds numbers}
In this section, we analyse the effect of including the corrective forcing in the TGV simulations. We study three Reynolds numbers, hyperparameters, and the usefulness of the forcing to correct the solution in different stages of the simulation (laminar, transitional, and turbulent). 

\begin{figure*}
 \centering
 \subfloat[\centering Mesh.]{
    \includegraphics[width=0.25\textwidth]{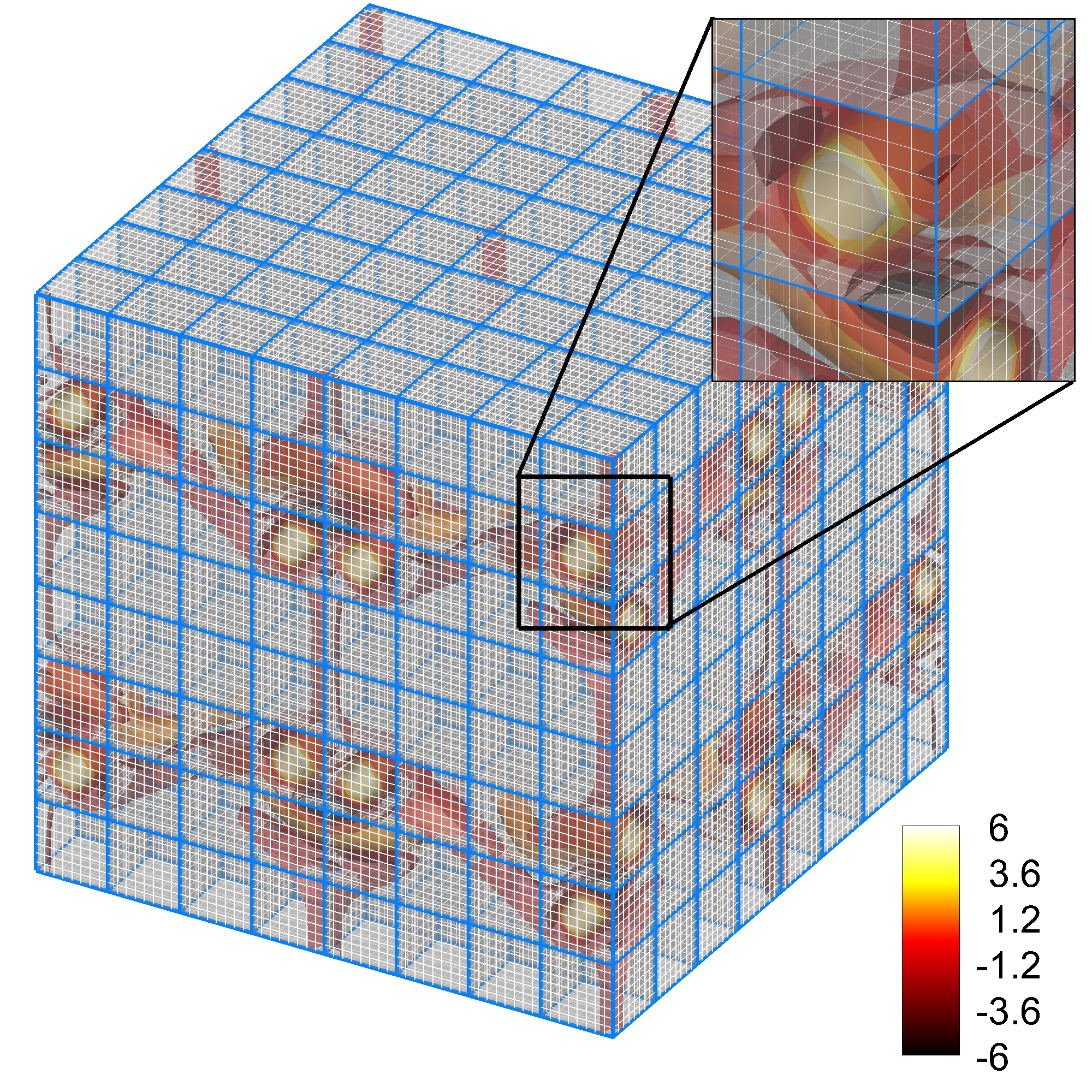}
     \label{Fig:BoxMesh}}
  \subfloat[\centering $Re =$ 30.]{
    \includegraphics[width=0.25\textwidth]{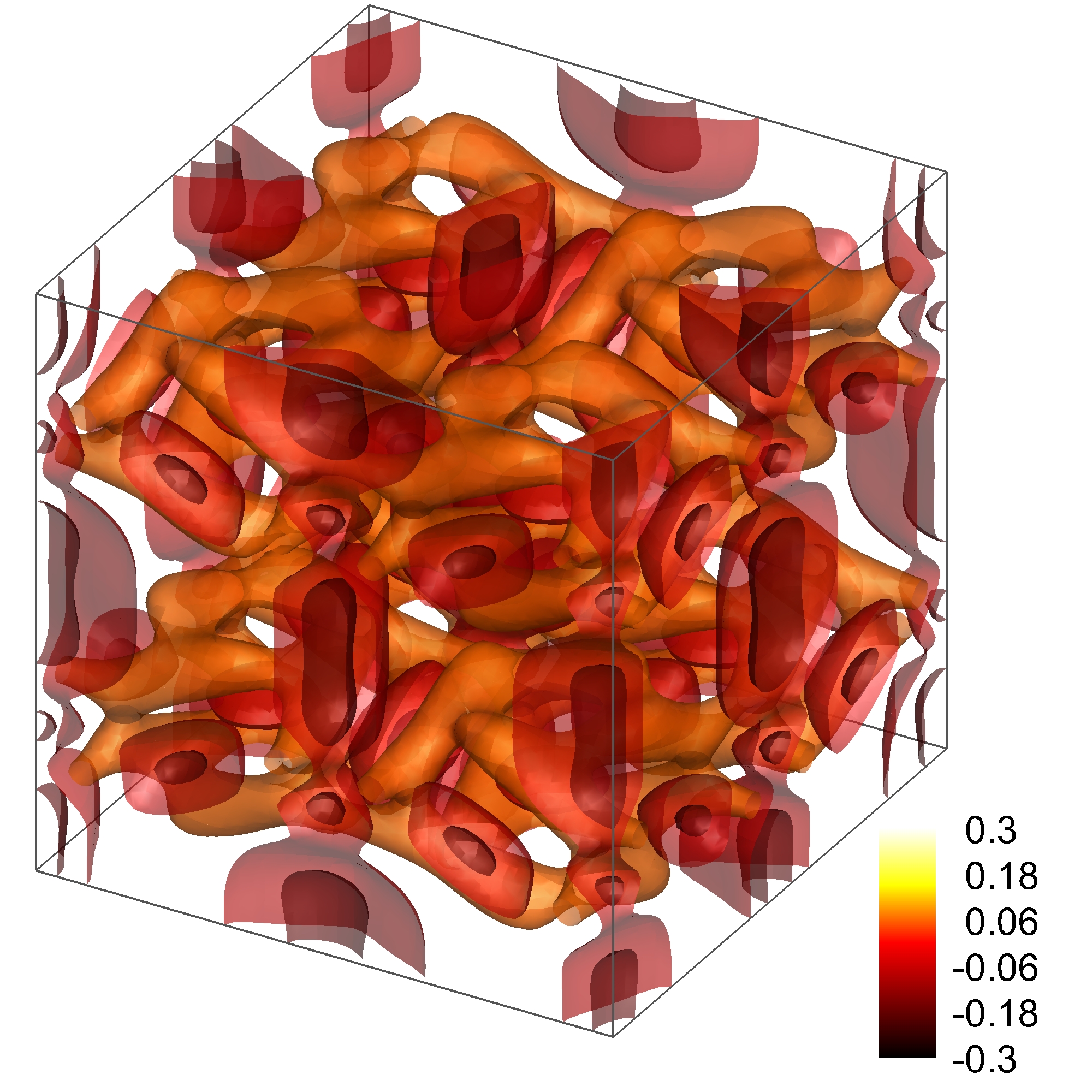}
     \label{Fig:IsoSurfaceRe30}}
  \subfloat[\centering $Re =$ 200.]{
    \includegraphics[width=0.25\textwidth]{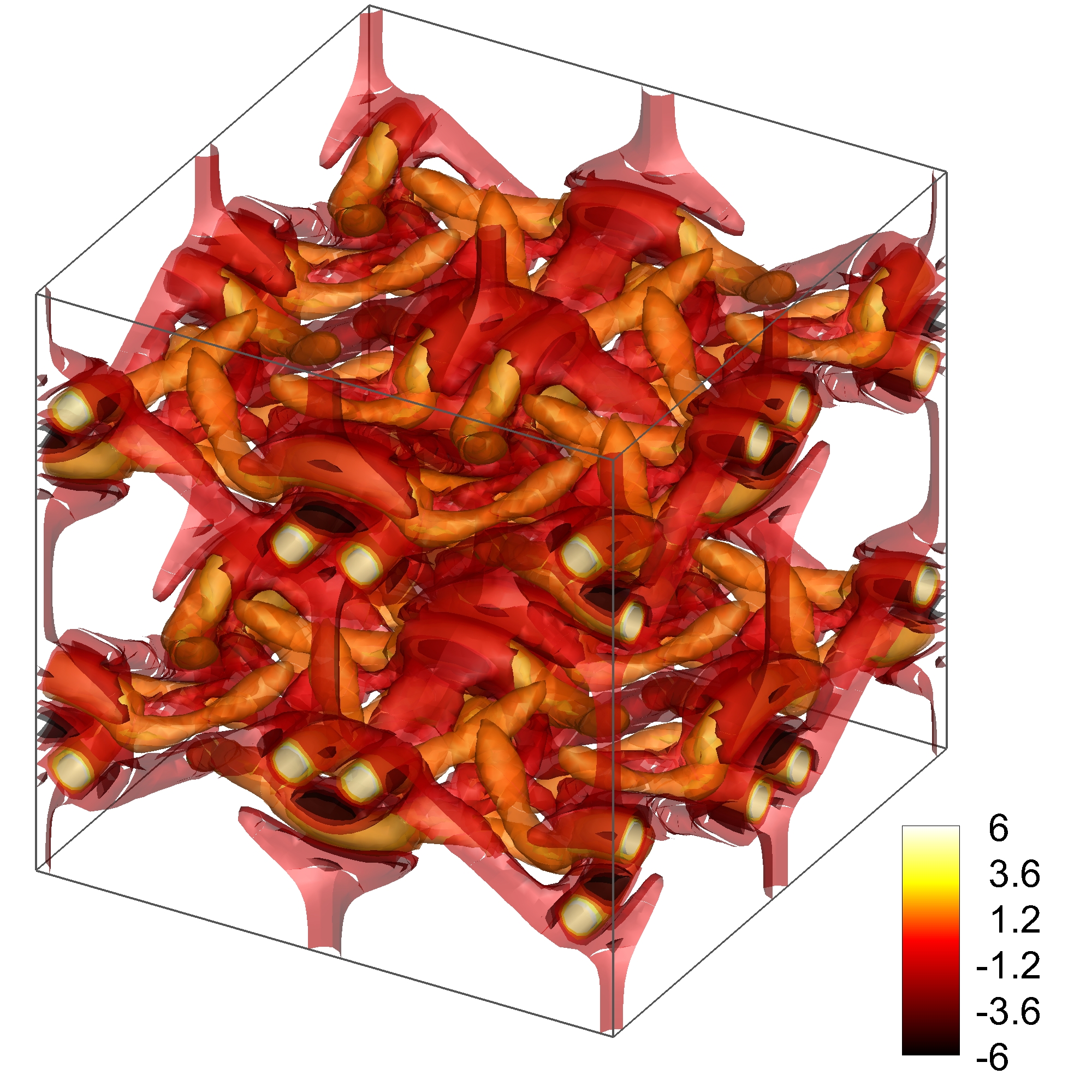}
    \label{Fig:IsoSurfaceRe200}}
   \subfloat[\centering $Re =$ 1600 LES.]{
    \includegraphics[width=0.25\textwidth]{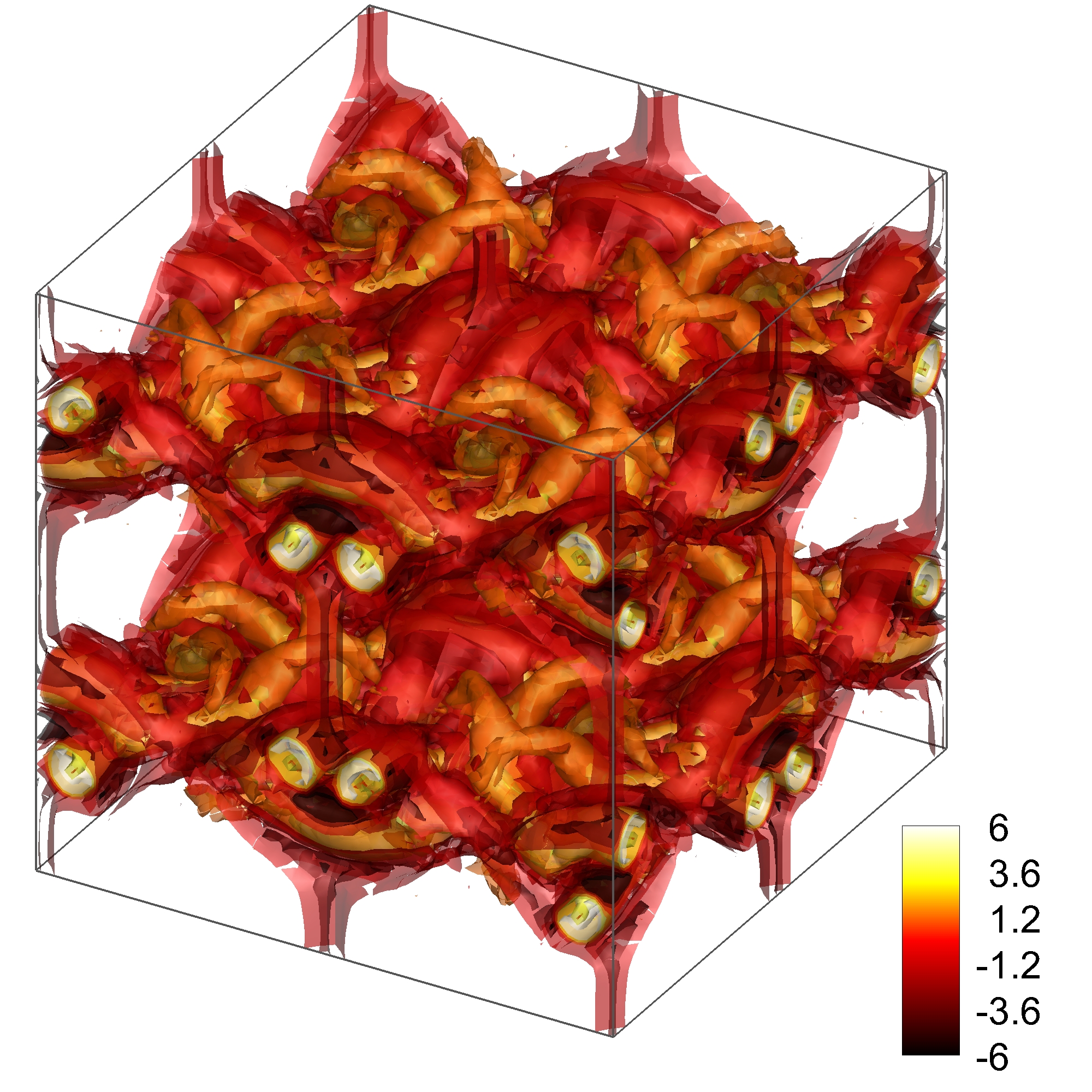}
    \label{Fig:IsoSurfaceRe1600_LES}}
    \caption{Q-criterion for different Reynolds numbers $Re=$30, 200 and 1600, at $t=$ 7 s}
    \label{Fig:DifferentReynodlsIsoSurface}
\end{figure*}

A summary of cases is provided in Table \ref{tab:CasesTable}, where we can see three cases for three Reynolds numbers.
\begin{itemize}
    \item Case 1: Reynolds number $Re=30$
    \item Case 2: Reynolds number $Re=200$
    \item Case 3: Reynolds number $Re=1600$
\end{itemize}
We have selected different Reynolds numbers to address the usefulness of the corrective forcing for smooth flows (low Re) and less smooth flows (high Re). Figure \ref{Fig:DifferentReynodlsIsoSurface} shows the velocity contours of the flow field in the three Reynolds numbers and illustrates the smoothness of the solution. High order methods provide spectral convergence (exponential decay of the error) for smooth flows \cite{kopriva2009implementing,Ferrer2010,LASKOWSKI2022110883,rubio2015truncation} and therefore showing that our corrective forcing can be useful for non-smooth flows is interesting. Note that we need to simulate a low polynomial order with the corrective forcing. 

In addition, note that we initialise the flow field with $C_{\infty}$ functions and, hence, at the start of the simulation, the flow is smooth but not isotropic. After some time ($t>6$ s) the flow generates small scales. Therefore, the flow is expected to change significantly during the first seconds of the simulation, and we could suspect that the neural network will not perform well outside the training region.
When the flow develops ($t>6$ s) the solution becomes isotropic and smaller scales develop. In these advanced times, the training data can be better extrapolated and used outside the training region (at least for a short time). To take this reasoning into account, in this section, we have tested the methodology for each Reynolds number at the starting conditions ($t=0$ s) and at advanced times ($t=7$ s). Note that in section \ref{sec:other_stages} we will test the methodology for Reynolds $Re=1600$ and additional time frames.

\begin{figure*}
 \centering
 \subfloat[\centering $t=$ 4 s.]{
    \includegraphics[width=0.25\textwidth]{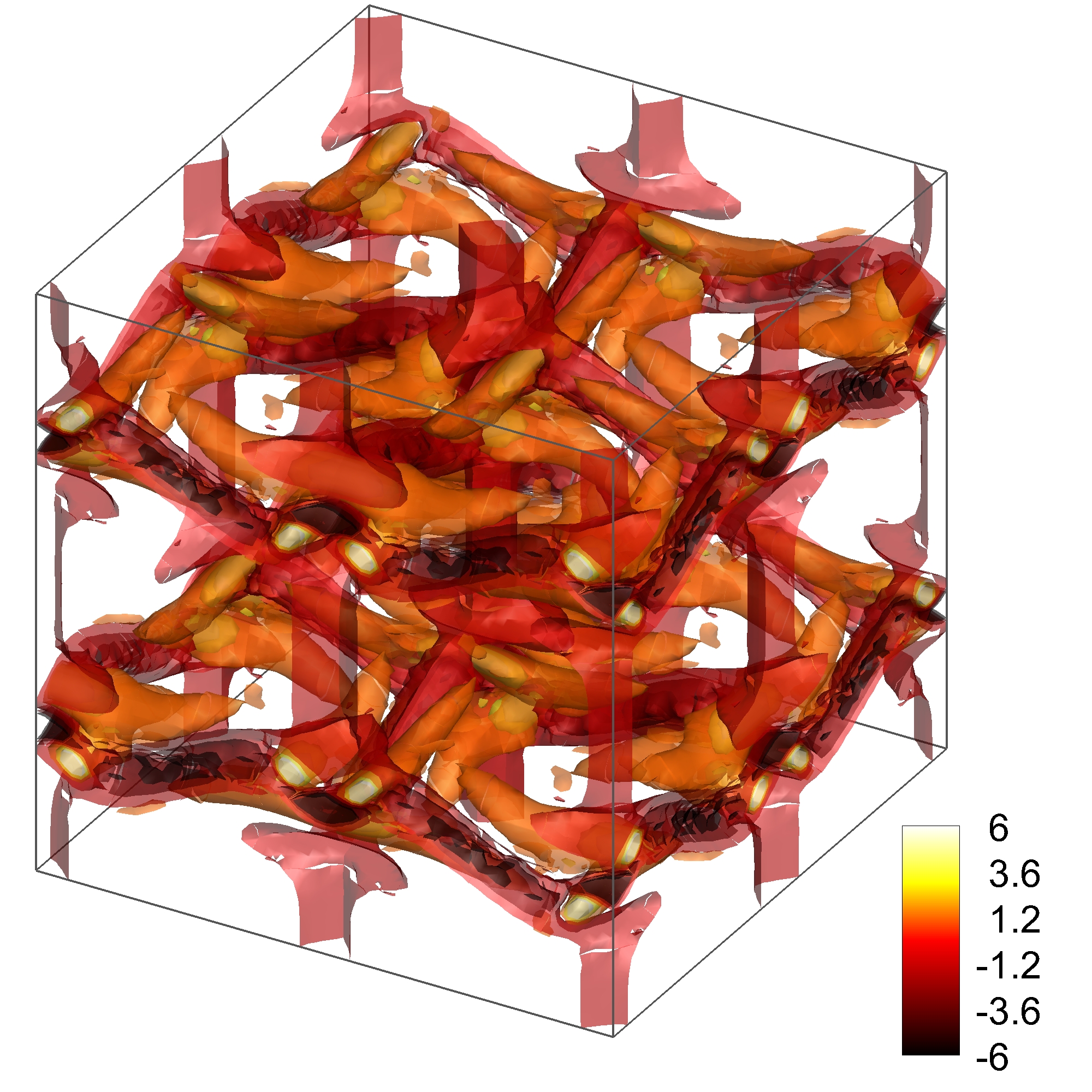}
     \label{Fig:IsoSurfaceLESt4}}
  \subfloat[\centering $t=$ 7 s.]{
    \includegraphics[width=0.25\textwidth]{images/LES_t7.png}
     \label{Fig:IsoSurfaceLESt7}}
  \subfloat[\centering $t=$ 10 s.]{
    \includegraphics[width=0.25\textwidth]{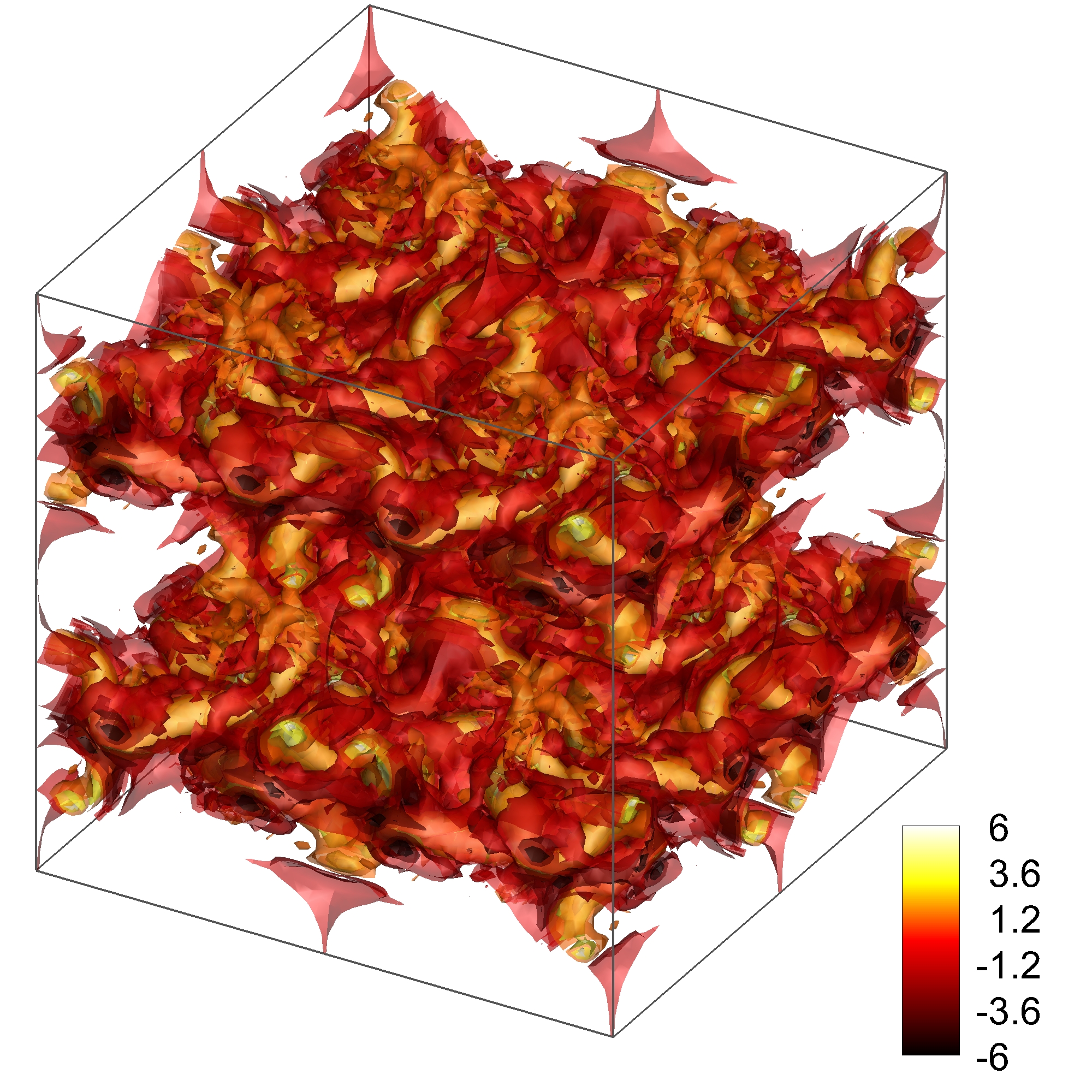}
    \label{Fig:IsoSurfaceLESt10}}
   \subfloat[\centering $t=$ 13 s.]{
    \includegraphics[width=0.25\textwidth]{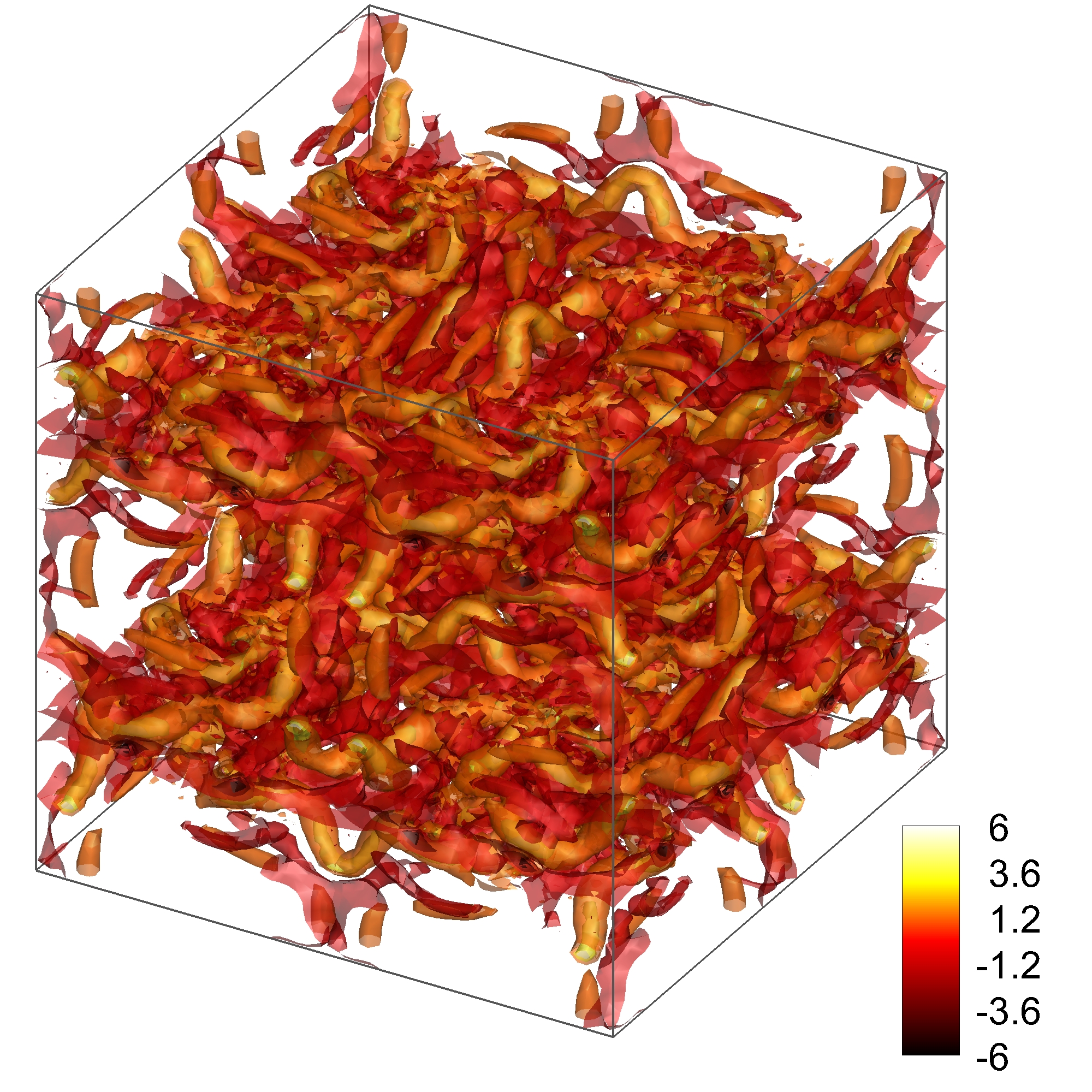}
    \label{Fig:IsoSurfaceLESt13}}
    \caption{Q-criterion for $Re=$1600 at different times $t=$4, 7, 10 and 13 s.}
    \label{Fig:LESDifferentTimes}
\end{figure*}

Each Reynolds number case is described in Table \ref{tab:CasesTable} and includes subcases: a) simulations at starting time $t=0$ s, b) and c) have advanced time $t=7$ s. In each subcase, we detail the set of equations used for the high order (HO scheme equations) and low order (LO scheme equations), including Navier-Stokes (NS), Large Eddy Simulation using Smagorinsky (LES), and NN when using the neural network corrective forcing. The polynomial orders for the high and low order solutions are detailed by High Order, $P_{HO}$ and Low Order, $P_{LO}$, respectively We also present the time steps for both the high order and low order schemes $\Delta t_{HO}$ and $\Delta t_{LO}$. Hyperparameters are also included: the number of layers, epochs, and batches for each case. The data used for training is in the interval given by the row: train time interval. From the results we can see the time necessary for the high order scheme to perform 400 low order time steps in column HO time 400 iter (S), and the time necessary for the low order scheme with and without corrective forcing to do 400 low order time steps in rows LO time 400 iter (S) and LO without neural network (s), respectively.  The ratio between the HO time cost and the LO time cost is presented in the row Ratio HO/LO. Finally, we include the time to train the neural network in the NN train time and the loss error achieved, measured as the mean square error.

\begin{table*}[]
\centering
\resizebox{\textwidth}{!}{%
\begin{tabular}{|c|cc|ccc|ccc|}
\hline
\begin{tabular}[c]{@{}c@{}}Case\\ Number\end{tabular}          & \multicolumn{2}{c|}{Case 1}                        & \multicolumn{3}{c|}{Case 2}                                                               & \multicolumn{3}{c|}{Case 3}                                                            \\ \hline
\begin{tabular}[c]{@{}c@{}}Reynolds\\ Number\end{tabular}      & \multicolumn{2}{c|}{$30$}                            & \multicolumn{3}{c|}{$200$}                                                                  & \multicolumn{3}{c|}{$1600$}                                                              \\ \hline
\begin{tabular}[c]{@{}c@{}} sub\\ case \end{tabular}                                                         & \multicolumn{1}{c|}{a}            & b              & \multicolumn{1}{c|}{a}            & \multicolumn{1}{c|}{b}              & c               & \multicolumn{1}{c|}{a}            & \multicolumn{1}{c|}{b}             & c             \\ \hline
\begin{tabular}[c]{@{}c@{}}HO scheme\\ equations \end{tabular}                                                         & \multicolumn{1}{c|}{NS}          & NS            & \multicolumn{1}{c|}{NS}          & \multicolumn{1}{c|}{NS}            & NS             & \multicolumn{1}{c|}{NS}          & \multicolumn{1}{c|}{LES}           & LES           \\ \hline
\begin{tabular}[c]{@{}c@{}}LO scheme\\ equations \end{tabular}                                                         & \multicolumn{1}{c|}{NS + NN}      & NS + NN        & \multicolumn{1}{c|}{NS + NN}      & \multicolumn{1}{c|}{NS + NN}        & NS + NN         & \multicolumn{1}{c|}{NS + NN}      & \multicolumn{1}{c|}{NS + NN}       & LES + NN      \\ \hline
\begin{tabular}[c]{@{}c@{}}Starting\\ time \end{tabular}                                                         & \multicolumn{1}{c|}{$0$}            & $7$             & \multicolumn{1}{c|}{$0$}            & \multicolumn{1}{c|}{$7$}             & $7$              & \multicolumn{1}{c|}{$0$}            & \multicolumn{1}{c|}{$7$}             & $7$             \\ \hline
\begin{tabular}[c]{@{}c@{}}Polynomial\\ order, $P_{HO}$ \end{tabular}                                                             & \multicolumn{1}{c|}{$8$}            & $8$              & \multicolumn{1}{c|}{$8$}            & \multicolumn{1}{c|}{$8$}              & $8$               & \multicolumn{1}{c|}{$8$}            & \multicolumn{1}{c|}{$8$}             & $8$             \\ \hline
\begin{tabular}[c]{@{}c@{}}Polynomial\\ order, $P_{LO}$ \end{tabular}                                                            & \multicolumn{1}{c|}{$2$}            & $2$              & \multicolumn{1}{c|}{$2$}            & \multicolumn{1}{c|}{$2$}              & $3$               & \multicolumn{1}{c|}{$2$}            & \multicolumn{1}{c|}{$3$}             & $3$             \\ \hline
\begin{tabular}[c]{@{}c@{}}time step\\ $\Delta t_{HO}$ \end{tabular}                                                         & \multicolumn{1}{c|}{$5 \cdot 10^{-4}$}    & $5 \cdot 10^{-4}$      & \multicolumn{1}{c|}{$5 \cdot 10^{-4}$}    & \multicolumn{1}{c|}{$5 \cdot 10^{-4}$}      & $5 \cdot 10^{-4}$       & \multicolumn{1}{c|}{$5 \cdot 10^{-4}$}    & \multicolumn{1}{c|}{$5 \cdot 10^{-4}$}     & $5 \cdot 10^{-4}$     \\ \hline
\begin{tabular}[c]{@{}c@{}}time step\\ $\Delta t_{LO}$ \end{tabular}                                                        & \multicolumn{1}{c|}{$3 \cdot 10^{-3}$}    & $3 \cdot 10^{-3}$      & \multicolumn{1}{c|}{$3 \cdot 10^{-3}$}    & \multicolumn{1}{c|}{$3 \cdot 10^{-3}$}      & $1.5 \cdot 10^{-3}$     & \multicolumn{1}{c|}{$3 \cdot 10^{-3}$}    & \multicolumn{1}{c|}{$1.5 \cdot 10^{-3}$}   & $1.5 \cdot 10^{-3}$   \\ \hline
\begin{tabular}[c]{@{}c@{}}Number\\ of Layers\end{tabular}                                                       & \multicolumn{1}{c|}{$4$}            & $4$              & \multicolumn{1}{c|}{$4$}            & \multicolumn{1}{c|}{$4$}              & $4$               & \multicolumn{1}{c|}{$4$}            & \multicolumn{1}{c|}{$5$}             & $5$             \\ \hline
\begin{tabular}[c]{@{}c@{}}Number\\ of epochs \end{tabular}                                                        & \multicolumn{1}{c|}{$30$}           & $30$             & \multicolumn{1}{c|}{$30$}           & \multicolumn{1}{c|}{$30$}             & $30$              & \multicolumn{1}{c|}{$30$}           & \multicolumn{1}{c|}{$35$}            & $50$           \\ \hline
\begin{tabular}[c]{@{}c@{}}Number\\ of batches\end{tabular}                                                        & \multicolumn{1}{c|}{$20$}           & $20$             & \multicolumn{1}{c|}{$20$}           & \multicolumn{1}{c|}{$20$}             & $20$              & \multicolumn{1}{c|}{$20$}           & \multicolumn{1}{c|}{$20$}            & $20$            \\ \hline
\begin{tabular}[c]{@{}c@{}}train time \\ interval (s)\end{tabular}       & \multicolumn{1}{c|}{{$[0,\:0.3]$}} & {$[7,\:7.3]$} & \multicolumn{1}{c|}{{$[0,\:0.3]$}} & \multicolumn{1}{c|}{{$[7,\:7.3]$}} & {$[7,\:7.15]$} & \multicolumn{1}{c|}{{$[0,\:0.3]$}} & \multicolumn{1}{c|}{{$[7,\:7.15]$}} & {$[7,\:7.15]$} \\ \hline
\begin{tabular}[c]{@{}c@{}}HO time\\ 400 iter (s)\end{tabular} & \multicolumn{1}{c|}{$1176.62$}      & $1443.83$        & \multicolumn{1}{c|}{$1351.88$}      & \multicolumn{1}{c|}{$1273.51$}        & $601.08$      & \multicolumn{1}{c|}{$1393.66$}      & \multicolumn{1}{c|}{$725.02$}        & $741.94$        \\ \hline
\begin{tabular}[c]{@{}c@{}}LO time\\ 400 iter (s)\end{tabular} & \multicolumn{1}{c|}{$18.04$}        & $29.57$          & \multicolumn{1}{c|}{$35.34$}        & \multicolumn{1}{c|}{$23.58$}          & $50.84$          & \multicolumn{1}{c|}{$26.45$}        & \multicolumn{1}{c|}{$43.65$}         & $59.49$         \\ \hline
\begin{tabular}[c]{@{}c@{}}Ratio\\ HO/LO\end{tabular}          & \multicolumn{1}{c|}{$65$}           & $49$             & \multicolumn{1}{c|}{$38$}           & \multicolumn{1}{c|}{$54$}             & $11.8$            & \multicolumn{1}{c|}{$53$}           & \multicolumn{1}{c|}{$16.6$}          & $12.5$          \\ \hline
\begin{tabular}[c]{@{}c@{}}LO no \\ nn (s)\end{tabular}        & \multicolumn{1}{c|}{$14.5$}         & $20.89$          & \multicolumn{1}{c|}{$21.83$}        & \multicolumn{1}{c|}{$20.63$}          & $31.35$           & \multicolumn{1}{c|}{$17.91$}        & \multicolumn{1}{c|}{$26.94$}         & $34.29$         \\ \hline
\begin{tabular}[c]{@{}c@{}}NN train\\ time (s)\end{tabular}    & \multicolumn{1}{c|}{$163.75$}       & $181.39$         & \multicolumn{1}{c|}{$171.36$}       & \multicolumn{1}{c|}{$175.60$}         & $317.36$          & \multicolumn{1}{c|}{$178.67$}       & \multicolumn{1}{c|}{$435.64$}        & $829.10$        \\ \hline
\begin{tabular}[c]{@{}c@{}}Loss\\ Error\end{tabular}           & \multicolumn{1}{c|}{$8.98 \cdot 10^{-5}$}   & $1.07 \cdot 10^{-4}$     & \multicolumn{1}{c|}{$8.9033 \cdot 10^{-5}$}   & \multicolumn{1}{c|}{$9.72 \cdot 10^{-5}$}     & $7.31 \cdot 10^{-5}$      & \multicolumn{1}{c|}{$8.95 \cdot 10^{-5}$}   & \multicolumn{1}{c|}{$1.09 \cdot 10^{-4}$}    & $1.60 \cdot 10^{-4}$    \\ \hline
\end{tabular}%
}
\caption{Summary of the test cases: Case 1, 2 and 3 include Reynolds numbers of $30$, $200$ and $1600$, respectively. The description of the subcases and notation are detailed in the main text. }
\label{tab:CasesTable}
\end{table*}

\subsection{Results for $Re=$30, 200 and 1600 at $t=$0 and 7 s }\label{sec:results_RE}
For each Reynolds number (case 1, case 2 and case 3), we test the methodology at the start ($t=0$ s) and at the middle of the simulation ($t=7$ s). 

\begin{figure*}
\centering
\subfloat[\centering Case1.a: $Re=30$.]{
\includegraphics[width = \textwidth]{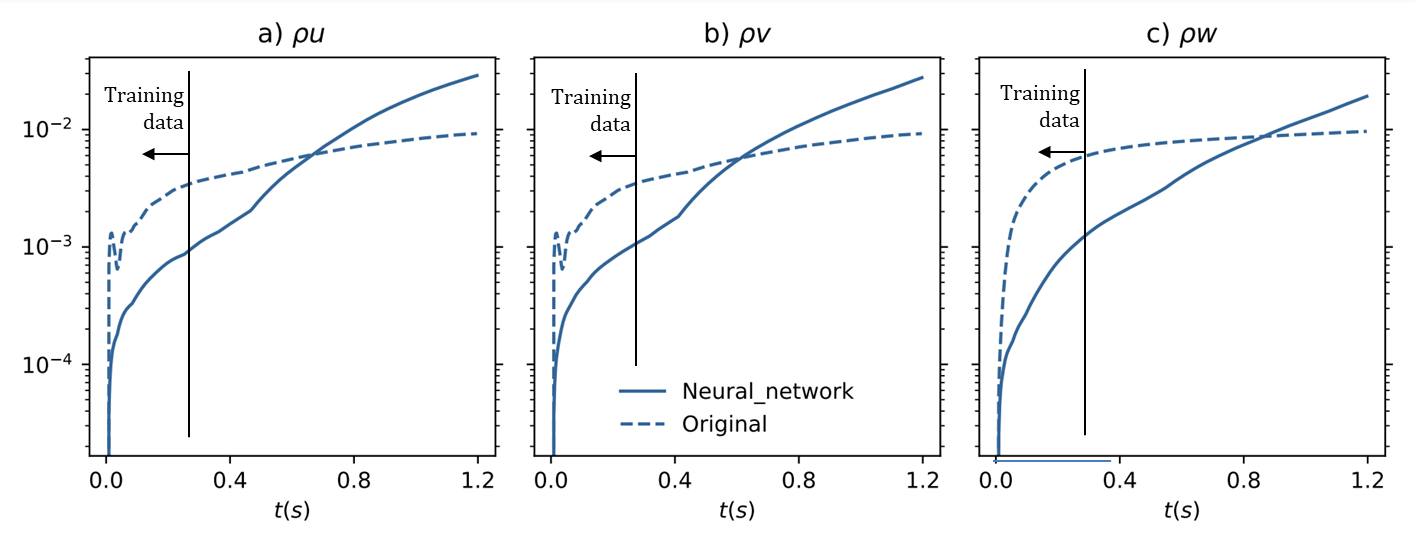}
\label{Case1a_error_comparison}}\\
\subfloat[\centering Case2.a: $Re=200$.]{
\includegraphics[width = \textwidth]{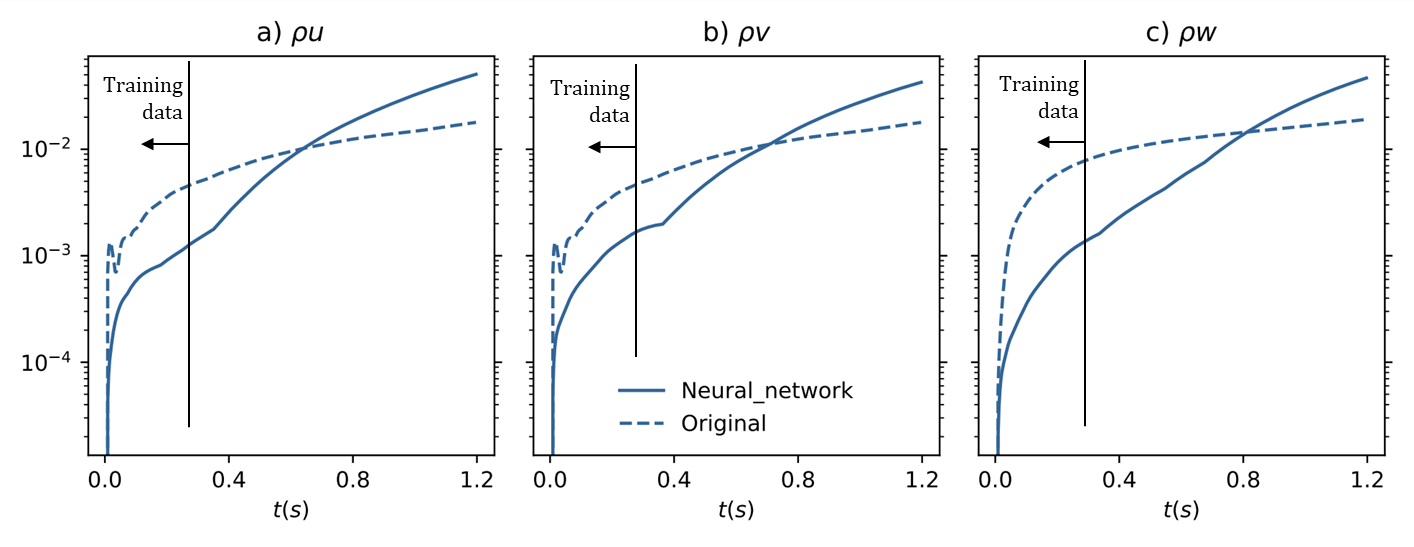}
\label{Case2a_error_comparison}}\\
\subfloat[\centering Case3.a: $Re=1600$.]{
\includegraphics[width = \textwidth]{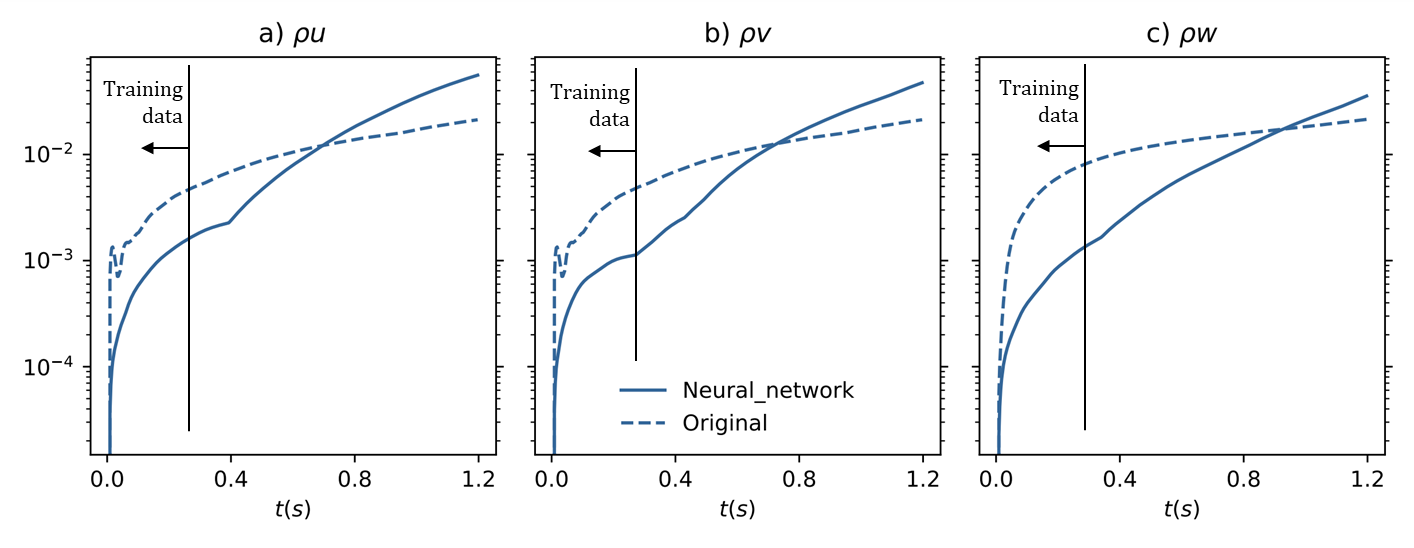}
\label{Case3a_error_comparison}}
\caption{Cases 1.a 2a and 3.a. with starting time $t=0$, showing infinite error in $\rho u$ (left), $\rho v$ (centre), and $\rho w$ (right). In all cases $P_{HO}=8$ to $P_{LO}=2$ and the training windows include $t=0$ to $t=0.3$ s.}
\end{figure*}

Figures \ref{Case1a_error_comparison}, \ref{Case2a_error_comparison} and \ref{Case3a_error_comparison}, show the results for $Re=$ 30, 200, and 1600 at the start ($t=0$ s). We train for a short time window of 0.3 seconds and see that the corrective forcing is only capable of keeping a lower error for another 0.3 seconds. In this case, the corrective forcing, trained for $t<t_{train}$, cannot be efficiently extrapolated to later times $t>t_{train}$ and confirms our hypothesis: during the start, the physics changes significantly (generating new flow scales), and the extrapolation from one window in time to another is not efficient. 


\begin{figure*}
    \centering
    \includegraphics[width = \textwidth]{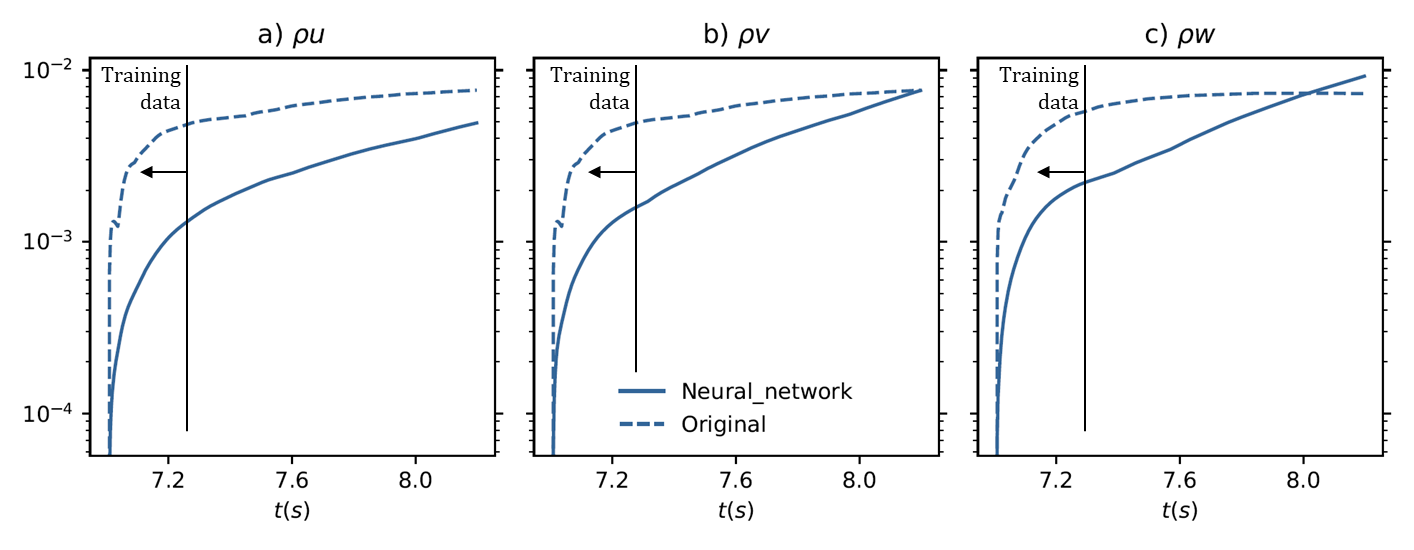}
    \caption{Case1.b: Infinite error in $\rho u$, $\rho v$ and $\rho w$, $Re=30$ and transformation from $P_{HO}=8$ to $P_{LO}=2$ with a starting time $t=7$ s. Trained from $t=7$ to $t=7.3$ s.}
    \label{Case1b_error_comparison}
\end{figure*}

Figure \ref{Case1b_error_comparison} shows case1.b ($Re=$ 30), which is trained in the middle of the simulation ($t=7$ s). Here, we obtain encouraging results since the corrective forcing allows us to decrease the error for another approximately 0.6 seconds (while the training window included 0.3s). In this case, we can check the cost per simulation to see if we obtain accelerations. We measure the time for $400$ iterations with the high order scheme and the low order scheme to see that the cost of high order was $65$ times larger than for the neural network case, see Table \ref{tab:CasesTable}. 

\begin{figure*}
\centering
\subfloat[\centering Case2.b: $P_{HO}=8$ to $P_{LO}=2$ and trained from $t=7$ to $t=7.3$ s.]{
\includegraphics[width = \textwidth]{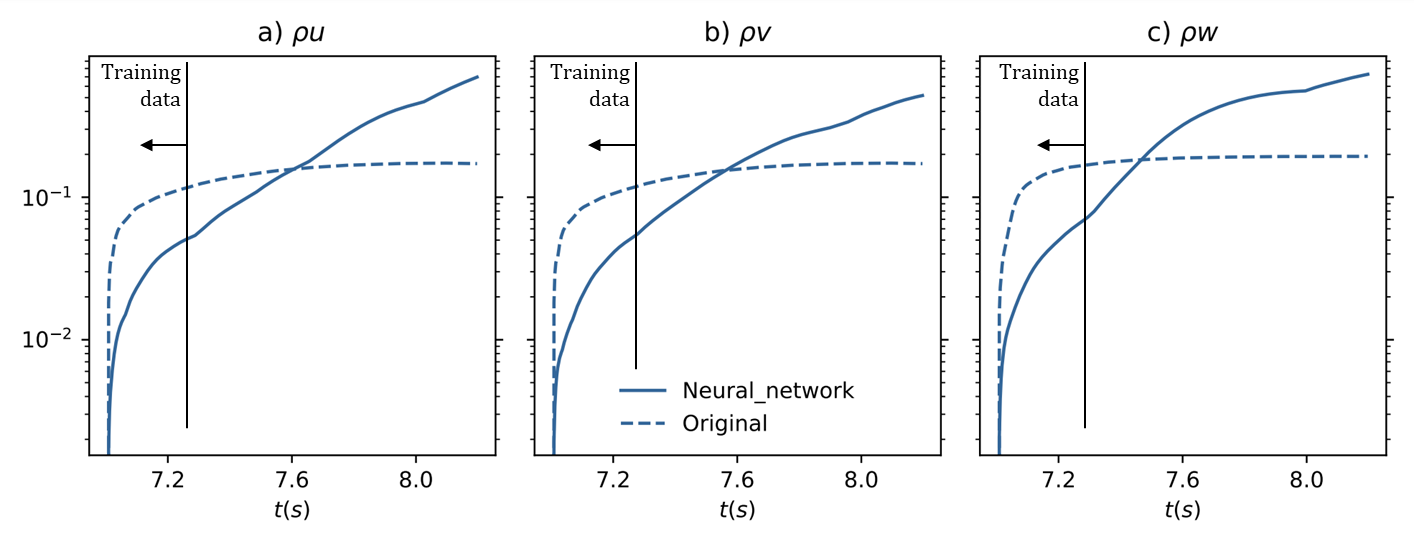}
\label{Case2b_error_comparison}}\\
\subfloat[\centering Case2.c: $P_{HO}=8$ to $P_{LO}=3$ and trained from $t=7$ to $t=7.15$ s.]{
\includegraphics[width = \textwidth]{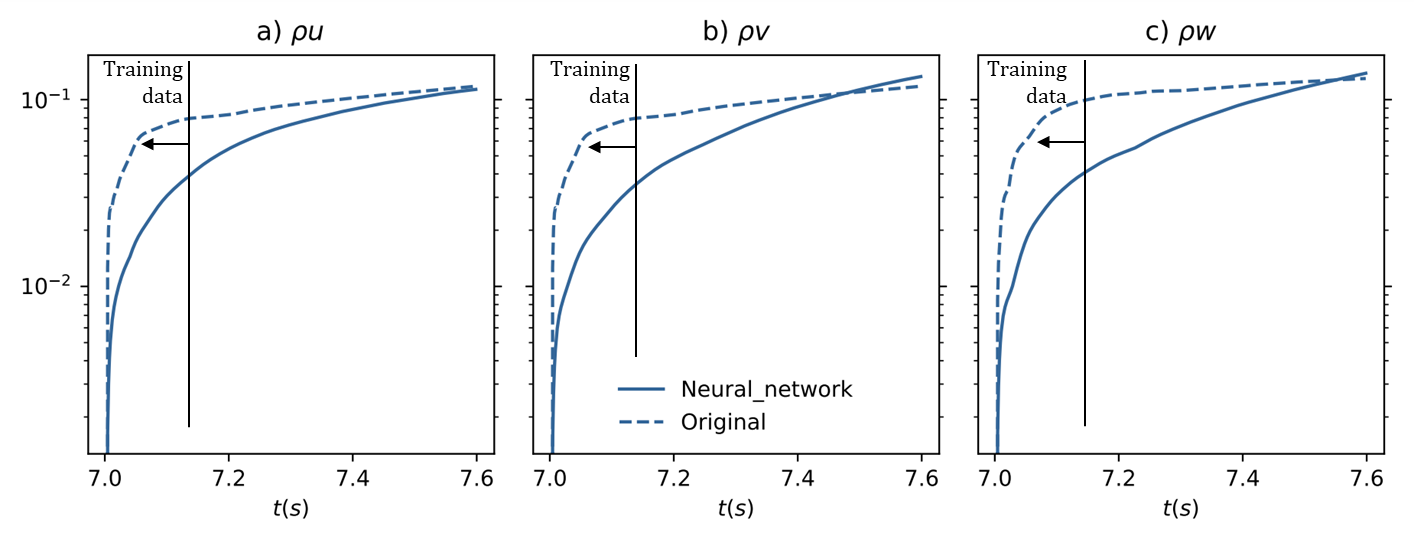}
\label{Case2c_error_comparison}}\\
\caption{Cases 2.b and 2.c. $Re=$200 with starting time $t=7$ s, showing an infinite error in $\rho u$ (left), $\rho v$ (centre), and $\rho w$ (right). }
\label{Case2a_error_comparison22}
\end{figure*}

Encouraged by the previous results, we try $Re=$200 at $t=7$, see Figure \ref{Case2b_error_comparison}, to observe bad results (we cannot use the trained corrective forcing efficiently after the training window).
Indeed, for $Re=$200, the flow is more complex (with a wider variety of scales) than for $Re=$30, see Figure \ref{Fig:DifferentReynodlsIsoSurface} and seems to be too demanding for the corrective forcing.
To overcome this difficulty, we increase the low polynomial order to $P_{LO}=3$. This small change in the low polynomial order  improves the solution significantly so that we can train the corrective forcing only for 0.15 seconds to obtain improvements for 0.6 seconds, see Figure \ref{Case2c_error_comparison}.

\begin{figure*}
\centering
\subfloat[\centering Case3.b: Only the high order solution has a LES model.]{
\includegraphics[width = \textwidth]{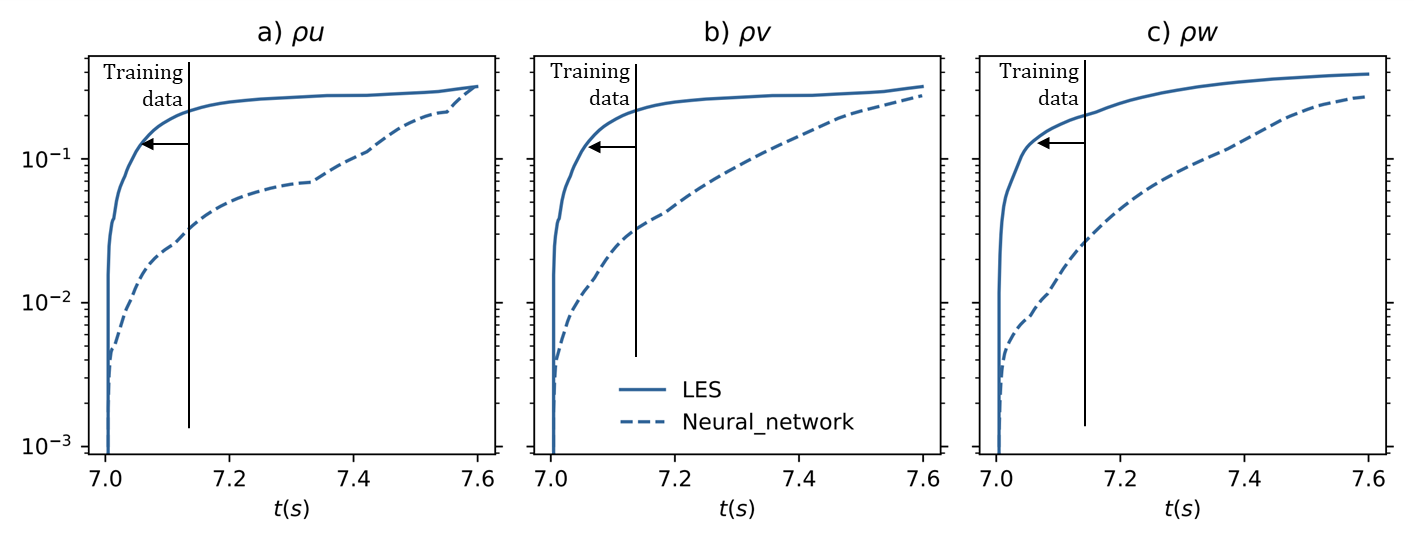}
\label{Case3b_error_comparison}}\\
\subfloat[\centering Case3.c: Both the high order and low order solutions have a LES model.]{
\includegraphics[width = \textwidth]{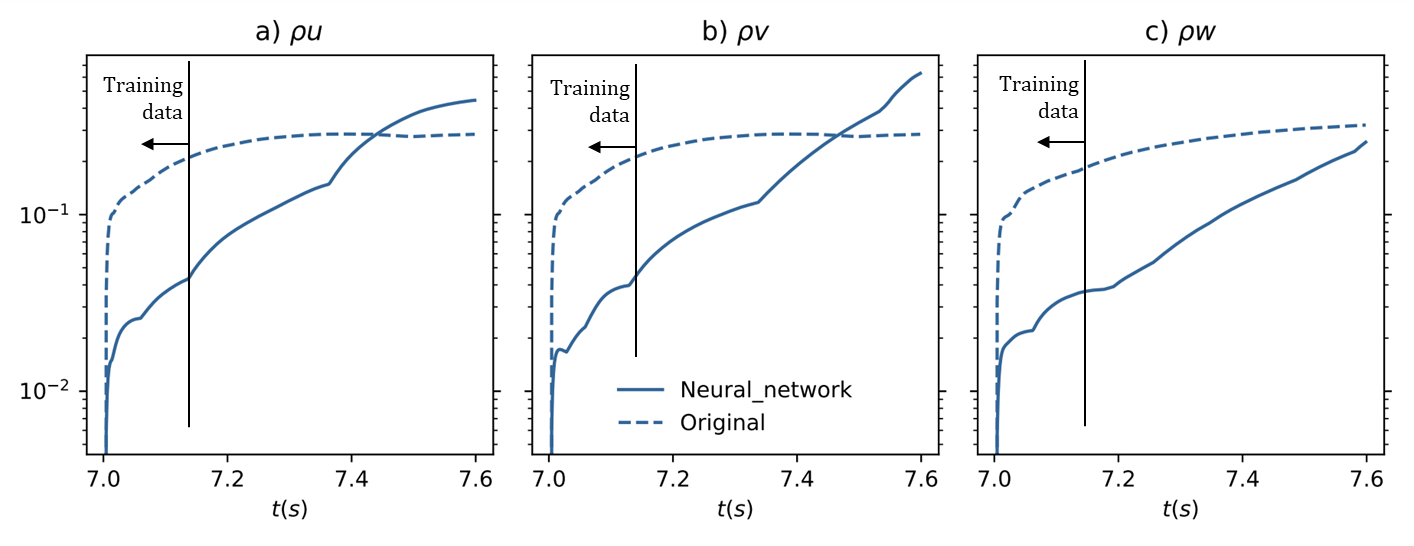}
\label{Case3c_error_comparison}}\\
\caption{Cases 3.b and 3.c. $Re=$1600 with starting time $t=7$ s, showing infinite error in $\rho u$ (left), $\rho v$ (centre), and $\rho w$ (right). In all cases $P_{HO}=8$ to $P_{LO}=3$ and the training windows include $t=7$ to $t=7.15$ s. }
\label{Case3b_error_comparison22}
\end{figure*}

Finally, we simulate $Re=$1600. In the high order solution, uses $P_{HO}=8$ and includes a Smagorinsky LES scheme. This allows us to study how general our methodology is as it can correct low order solutions with and without turbulent models. 
 Regarding the low  polynomial orders with corrective forcing, we tested two cases for $t=7$ s: case3.b which does not use a Smagorinsky LES model (Figure \ref{Case3b_error_comparison}) and case3.c where we include the corrective forcing in addition to the Smagorinsky LES model (Figure \ref{Case3c_error_comparison}). Comparing these alternatives in Figure \ref{Case3b_error_comparison22}, we can see that it is better to exclude the Smagorinsky LES model from the low order solution and allow the corrective forcing to account for the subgrid model. This suggests that NN has the potential to be used as a subgrid scale model, as also shown by Beck \textit{et al.} \cite{BECK2019108910}. However, let us note that our low order simulation is very coarse and hence our corrective forcing is doing more that modelling the effect of small scales, showing that our methodology is not focused on turbulence modelling but rather on accounting for all missing scales in the flow (independently of their size).



We conclude that the proposed methodology can improve the accuracy of the simulation for all Reynolds numbers considered. We have noted that the corrective forcing is not very useful for very in-homogeneous flows ($t=$0 s) but that once the flow has reached a certain degree of homogeneity, we obtain accurate results. To further explore this last point, we analyse the errors for various times (and $Re=$1600).

\subsection{Results for various time frames (flow regimes) at $Re=$1600}\label{sec:other_stages}
\begin{figure*}
\centering
\subfloat[\centering $t=$4 s.]{
\includegraphics[width = \textwidth]{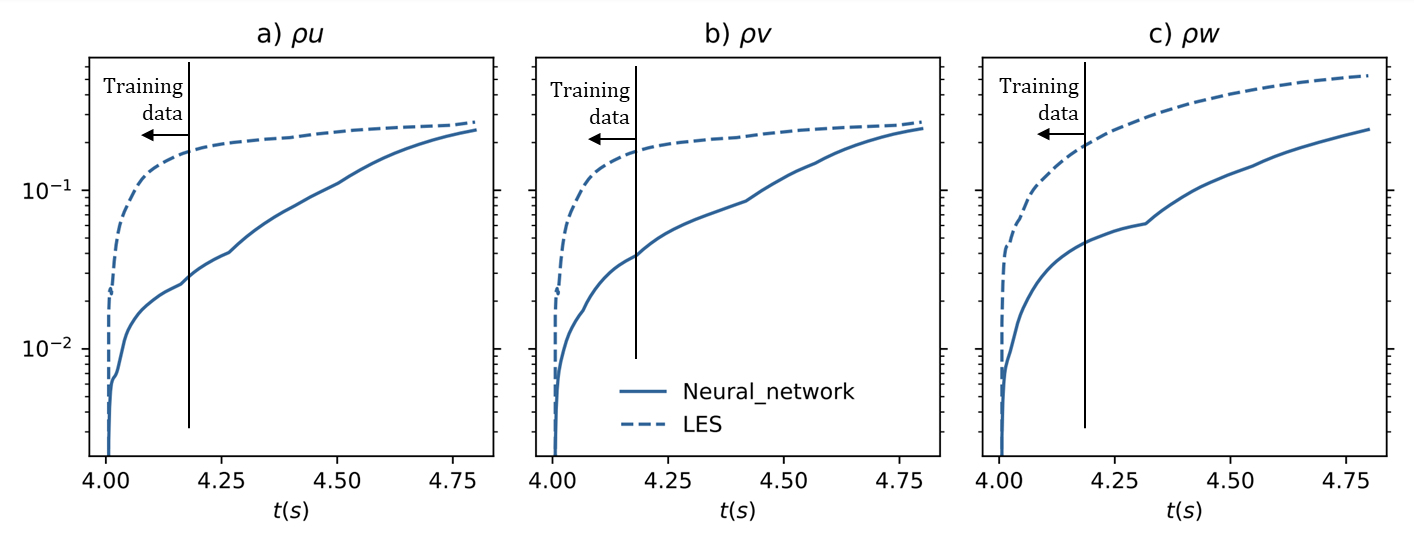}
\label{Case_T4_error_comparison}}\\
\subfloat[\centering $t=$10 s.]{
    \includegraphics[width=\textwidth]{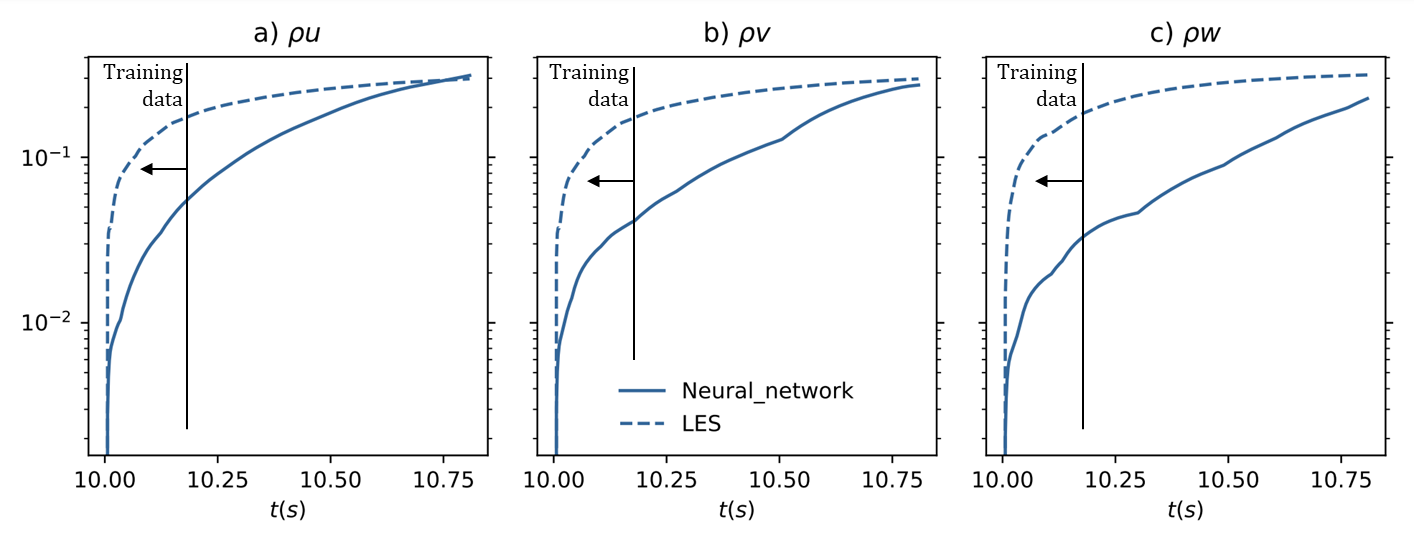} 
    \label{Subfig:Case_T10_error_comparison}} \\
\subfloat[\centering $t=$13 s.]{
    \includegraphics[width=\textwidth]{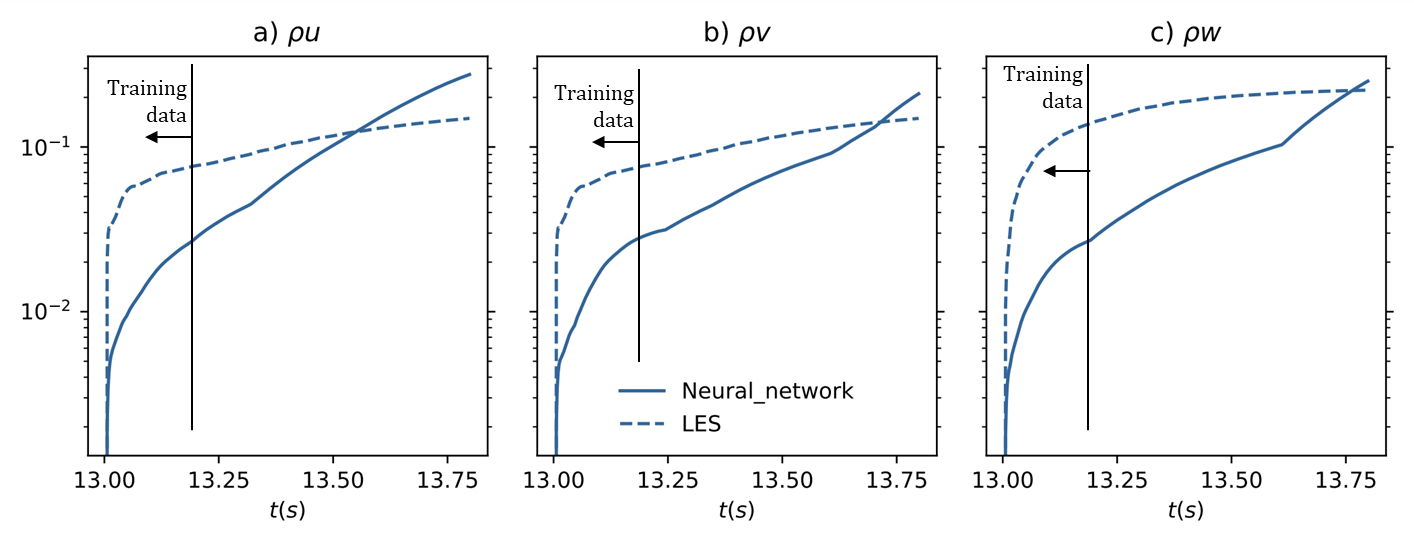} 
    \label{Subfig:Case_T13_error_comparison}}
\caption{$Re=$1600 for 3 different simulation times: $t=$4, 10 and 13 s, showing infinite error in $\rho u$ (left), $\rho v$ (centre), and $\rho w$ (right). In all cases $P_{HO}=8$ and $P_{LO}=3$. }
\label{varios_tiempos}
\end{figure*}

In the TGV problem, the flow evolves considerable throughout the simulation, including laminar, transitional, and turbulent flows. In this section, we verify that our methodology is useful at different stages of the simulation, in addition to $t=$7 s, which was already considered in section \ref{sec:results_RE}. We consider $t=$ 4 s, $t=$ 10 s, and $t=$ 13 s. These three situations have been chosen because they are representative conditions in the TGV test case (see Figure \ref{Fig:LESDifferentTimes}). The NN architecture is unchanged in all cases and trained for 100 iterations (0.15 s) from the beginning of each time frame. The low order simulation with corrective forcing does not include an LES model. The error curves are shown in Figure \ref{varios_tiempos}. The errors obtained for $P=$3, including corrective forcing, show significantly lower errors than when forcing is not included and for a long time after training. 

In summary, the corrective forcing provides accuracy for the different flow regimes covered by the different times included in this section.

\subsection{Effective accuracy and acceleration}
The error achieved by the corrective forcing has been in multiple cases lower than the error given by the low order solution without forcing, showing the usefulness of the proposed methodology. However, we have not addressed the issue of accuracy achieved. In this section, we ask: What is the real accuracy of the corrected low-order method? and how much faster is our corrected solution for a given accuracy? 

\begin{figure*}
    \centering
    \includegraphics[width = \textwidth]{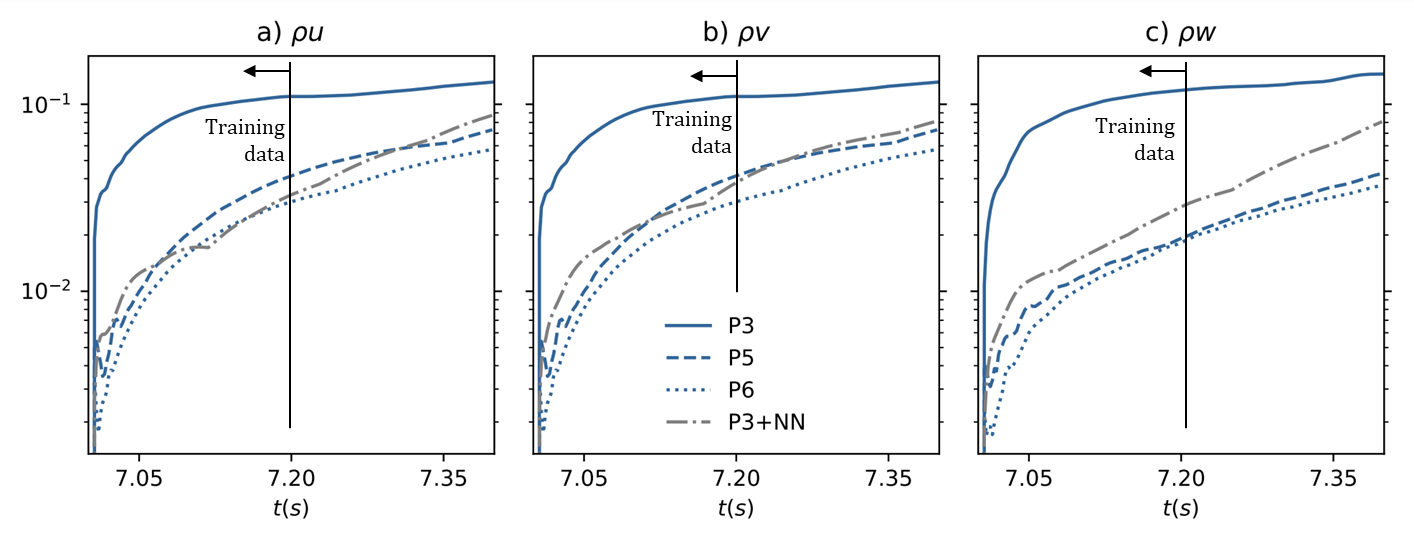}
    \caption{$Re=$1600 simulated with various polynomial orders: $P=$3,  $P=$3+NN, $P=$5 and $P=$6, showing infinite error in $\rho u$ (left), $\rho v$ (centre), and $\rho w$ (right). For the neural network case ($P=$3+NN) we have used 8 layers with 30 epochs and the corrective forcing was trained with data from $t=7$ s to $t=7.2$ s.}
    \label{EffectiveAcceleration_error_comparison}
\end{figure*}

To answer these questions, we consider $Re$=1600 at $t=$7 s (case 3b), with a high  polynomial order of $P_{HO}=8$ and a reduction to a low polynomial order of $P_{LO}=3$. We run additional simulations for polynomial orders $P=$5 and 6. We computed the errors with respect to the high polynomial order and depict the error curves in Figure \ref{EffectiveAcceleration_error_comparison}. It can be seen that the $P=$3 solution that includes the corrective forcing (P3+NN) provides comparable accuracy to the $P=$5 (P5) and $P=$6 (P6) solutions, and much better accuracy than the $P=$3 without forcing (P3).

In addition, in Table \ref{tab:my-table}, the time steps used, the wall clock time to simulate 200 iterations, and the ratio of wall clock times for all simulations, with respect to $P=3$ with the corrective forcing ($P_{LO}+NN$). As expected, the computational cost increases when the polynomial order increases. We can see that including the corrective forcing doubles the cost of simulating without forcing, but leads to an accuracy improvement of roughly 3 polynomial orders. Since the accuracy of $P_{LO}+NN$ is similar to $P=$5-6 (see Figure \ref{EffectiveAcceleration_error_comparison}), we can conclude that the proposed method is between $3.23$ and $4.94$ times faster, for this particular flow condition.  We conclude that the proposed methodology improves accuracy while achieving accelerations for 3D turbulent cases. 

We can conclude from this section that our methodology enables improved accuracy and provides faster computations, for an equivalent accuracy. 
\begin{table}[H]
\centering
\begin{tabular}{|c|c|c|c|c|}
\hline
Case                   & Order & $\Delta t$        & \begin{tabular}[c]{@{}c@{}}Cost (s)/\\ 200 iter\end{tabular} & Ratio \\ \hline
$P_{LO}$                  & 3     & $20 \cdot 10 ^{-4}$  & $13.47$                                                        & $0.50$  \\ \hline
$P_{LO}$ +NN                 & 3     & $20 \cdot 10 ^{-4}$  & $26.88$                                                        & $1$     \\ \hline
\multirow{3}{*}{$P_{IO}$ } & 4     & $10 \cdot 10 ^{-4}$  & $36.16$                                                        & $1.35$  \\ \cline{2-5} 
                       & 5     & $6.7 \cdot 10 ^{-4}$ & $86.94$                                                        & $3.23$  \\ \cline{2-5} 
                       & 6     & $6.7 \cdot 10 ^{-4}$ & $132.79$                                                       & $4.94$ \\ \hline
$P_{HO}$                   & 8     & $5.0 \cdot 10 ^{-4}$ & $532.07$                                                       & $19.79$ \\ \hline
\end{tabular}%
\caption{Effective acceleration for $Re$=1600 at $t=$7 s.}
\label{tab:my-table}
\end{table}


\subsection{Influence of the hyperparameters and \textit{transfer learning}}\label{sec:hyperparameters}
In this final section, we apply the methodology for $t=10$ s and evaluate the hyperparameters (layers and epochs) of the NN used to train the corrective forcing. Furthermore, we will study \textit{transfer learning} for a NN trained at $t=$0 s and used at $t=$7s, and also between Reynolds numbers $Re=$200 and $Re=$1600.



\begin{figure*}
    \centering
    \includegraphics[width = \textwidth]{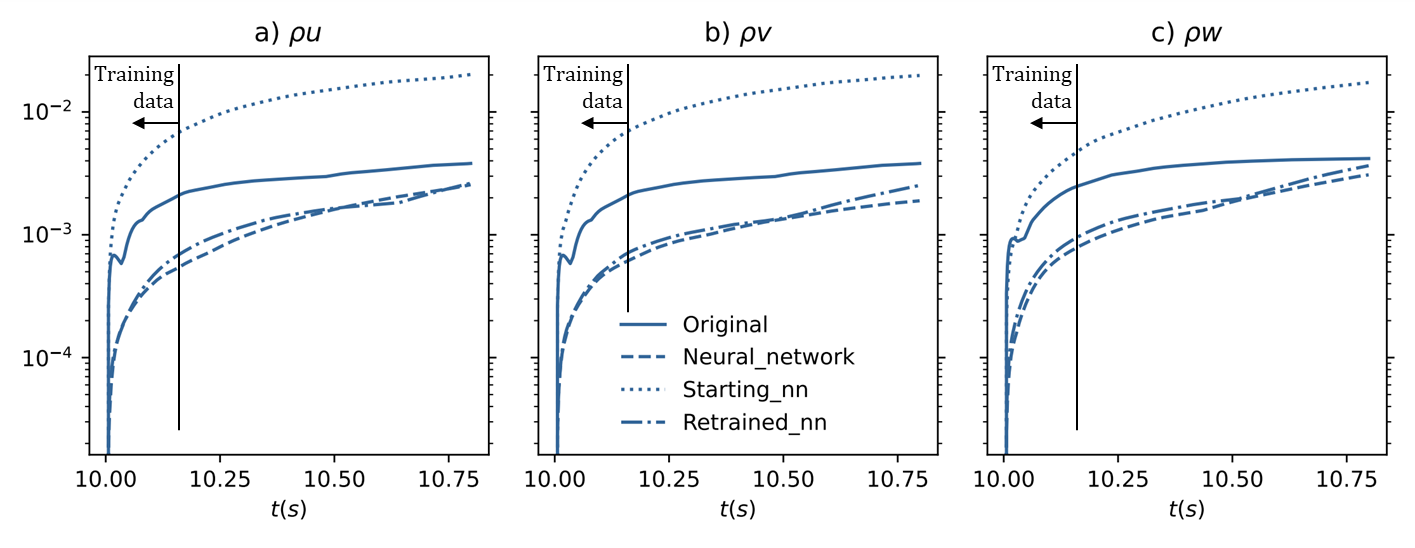}
    \caption{\textit{Transfer learning} between different time frames at $Re=30$ (case 1b), showing infinite error in $\rho u$ (left), $\rho v$ (centre), and $\rho w$ (right).  In all cases $P_{HO}=8$ to $P_{LO}=2$ and starting time is $t=10$ s.}
    \label{Case1b_test_error_comparison}
\end{figure*}

Figure \ref{Case1b_test_error_comparison} considers the case1.b ($Re=$30), where we increase the number of epochs and layers to show the potential for further improvement when considering more complex NNs to calculate the corrective forcing. We consider 4 cases: 
\begin{itemize}
    \item Original: Without neural network
    \item Neural network 1: 30 epochs and 4 layers (case 1b)
    \item Neural network 2: 40 epochs and 4 layers
    \item Neural network 3: 40 epochs and 5 layers
\end{itemize}
 We observe significant variations towards the end of the simulated window, where the most costly NN (40 epochs and 5 layers) provides lower errors. However, these effects are not significant, and benefits and accelerations are already seen when using 30 epochs and 4 layers (case 1b), which is the setting retained throughout this work.
\begin{figure*}
    \centering
    \includegraphics[width = \textwidth]{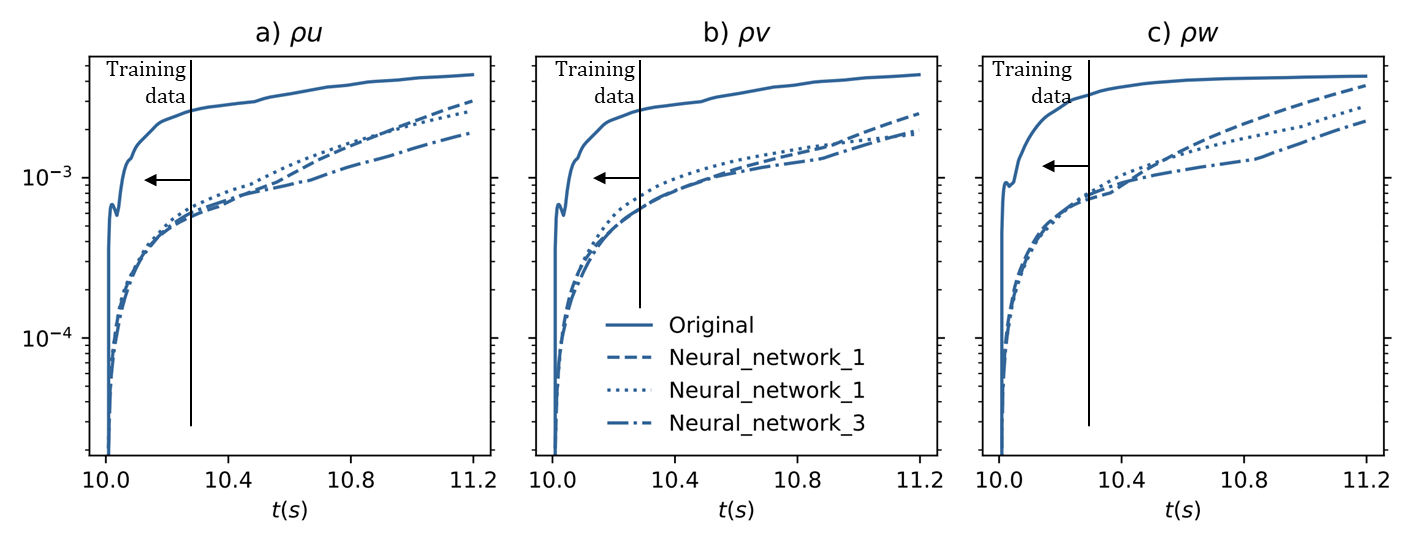}
    \caption{Sensitivity to the hyperparameters for $Re=$30 (case 1b with modifications), showing infinite error in $\rho u$ (left), $\rho v$ (centre), and $\rho w$ (right).  In all cases $P_{HO}=8$ to $P_{LO}=2$ and starting time is $t=10$ s.}
    \label{Case1b_variantes_error_comparison}
\end{figure*}



An interesting and popular possibility within the machine learning community is to perform \textit{transfer learning} \cite{vinuesa2022enhancing}, where an NN trained for a particular flow condition is then used to evaluate a corrective forcing for another flow. This technique minimises training costs, since one can train an NN only once and reuse it for other flow conditions. We use $Re=30$ to analyse this possibility by training a corrective forcing for $t=$0 s and using it to correct the low order solution at using a NN at $t=$10 s. This possibility is called (Starting\_nn) and the results are shown in Figure \ref{Case1b_test_error_comparison}. It can be seen that the errors are much larger than when not using any NN (Original), showing that for this case the flow changes significantly from $t=$0 s to $t=$10 s (see Figure \ref{Fig:DifferentReynodlsIsoSurface}) and therefore the corrective forcing is not useful at $t=$10 s. An alternative is to use the weights and bias (i.e., the corrective forcing) obtained when training at $t=$0 s, as the initial condition for the NN that we use at $t=$10 s, and retrain. We call this option (Retrained\_nn) and Figure \ref{Fig:DifferentReynodlsIsoSurface} shows that we obtain very similar results to when training directly at $t=$10 s (Neural\_network). However, when reusing the initial NN, we only need 10 epochs to find the ideal corrective forcing, which translates into spending 69.995 seconds of wall-clock computer time for training, instead of 181.39 seconds.

%

\begin{figure*}
    \centering
    \includegraphics[width = \textwidth]{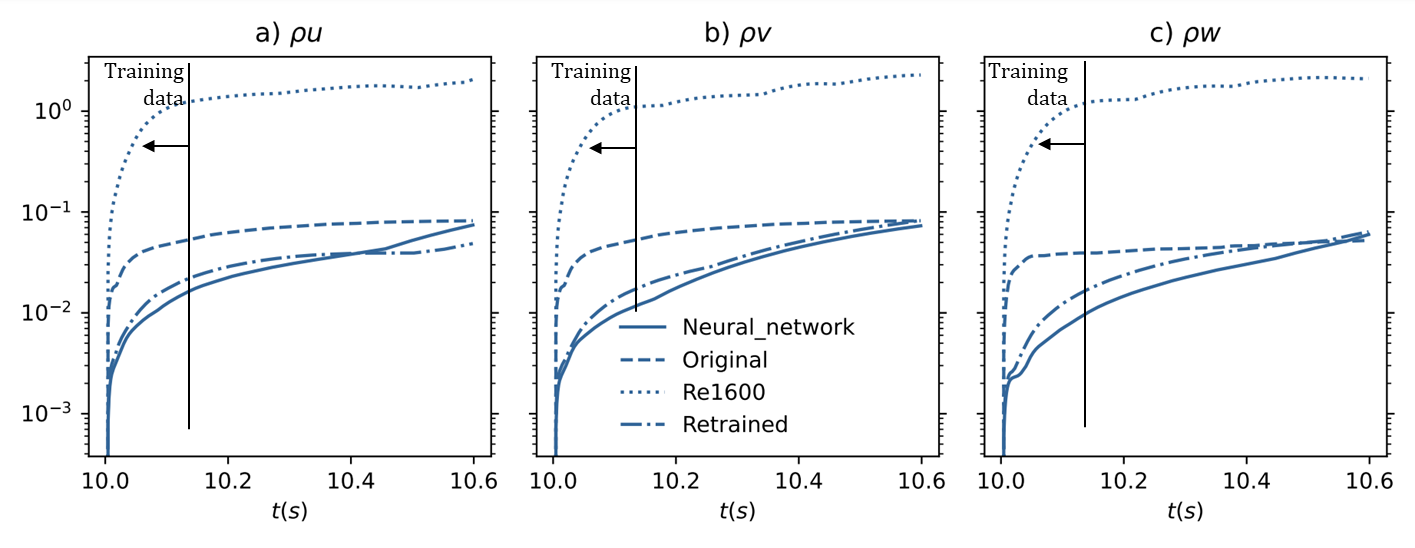}
    \caption{\textit{Transfer learning} between Reynolds numbers (for case 2c at $Re=200$),showing infinite error in $\rho u$ (left), $\rho v$ (centre), and $\rho w$ (right).  In all cases $P_{HO}=8$ to $P_{LO}=3$ and starting time is $t=10$ s.}
    \label{Case2c_test_error_comparison}
\end{figure*}

Finally, we explore the possibility of performing \textit{transfer learning} from $Re=$1600 (case 3b) to $Re=$200 (case 2c), since these flows are visually similar; see Figure \ref{Fig:DifferentReynodlsIsoSurface}. Figure \ref{Case2c_test_error_comparison} shows that using the corrective forcing trained for $Re=$1600 provides larger errors than when using no forcing (Original). However, once again 
 if we use the weights and bias (i.e., the corrective forcing), obtained for $Re=$1600, as an initial condition to optimise the neural network at $Re=$200 (Retrained), we obtain good results, which are comparable to the corrective forcing trained for $Re=$200. The re-trained NN converges in only 20 iterations, taking 227.00 seconds of wall-clock time for training, while the NN trained directly at $Re=$200 takes 440 seconds of wall-clock time. 

The results included in this section show that the selected hyperparameters are appropriate to provide an accurate forcing at reasonable cost, but of course these can be improved in further work. Furthermore, we have shown that \textit{transfer learning} cannot be applied directly, but that retraining an already trained forcing (under different flow conditions) can be very advantageous. 

\section{Conclusions}
\label{SecConclusions}
This paper is a continuation of \cite{DELARA2022105274}, where we presented simulations for the 1D Burgers equation. In this work, we extend the methodology for 3D turbulent flows governed by the Navier-Stokes equations with and without a Large Eddy Simulation closure model. The methodology evolves a solution with a low-order polynomial and includes a corrective forcing to obtain a more accurate solution. The corrective forcing is obtained using a fully connected deep NN. 

We have provided results for the 3D Taylor Green Vortex case, which includes laminar, transitional, and turbulent regimes as time progresses. We show the usefulness of the method in providing accurate results for various Reynolds values (30, 200 and 1600) and for different times (and flow regimes) during the simulation. In all cases, the methodology proved to be useful, as it was able to correct the solution outside the time frame considered for training. We have compared the accuracy provided by the methodology with simulations with other polynomial orders and showed an effective acceleration of 4-5.  

We analyse the effect of changes in hyperparameters and also explore \textit{transfer learning}. Regarding the former, we conclude that the hyperparameters need to be tuned for each flow condition, but that the selected ones provide reasonable accuracy for a variety of Reynolds and simulation times. Regarding the latter, the use of an NN trained with other flow conditions did not prove useful in our case. However,  using an already trained NN as the initial condition for a new training, we showed that we could obtain an appropriate corrective forcing only with few training iterations.   

In future work, the process of training, advancing a corrected low order scheme and retraining will be made automatic throughout a long simulation. For this, we will reuse an already trained corrective forcing thought \textit{transfer learning} and retrain it, for a shot time, once the errors become large. In addition, more sophisticated neural network architectures (e.g., convolutional or recurrent) will be explored, where scale interactions in space and time can be exploited to obtain a more accurate corrective forcing.

\newpage

\newpage
 
\appendix
\section{Compressible Navier-Stokes Equations}
\label{sec:cNS}
 
The 3D Navier-Stokes equations can be compactly written as

\begin{equation}
\partial_t \bm{q}  =- \nabla \cdot \blocktensor{f}^a + \nabla \cdot \blocktensor{f}^{\nu},
\label{NSeqs}
\end{equation}
where 
$\boldsymbol  q = [ \rho , \rho v_1 , \rho v_2 , \rho v_3 , \rho e]^T$ is the vector of the conservative variable, $\blocktensor{f}^a$ are the fluxes of the inviscid or Euler equations,

\begin{equation}
\blocktensor{f}^a = \left[\begin{array}{ccc} \rho v_1 & \rho v_2 & \rho u_3 \\
                                                                                \rho v_1^2 + p & \rho v_1v_2 & \rho v_1v_3 \\
                                                                                	\rho v_1v_2 & \rho v_2^2 + p & \rho v_2v_3 \\
                                                                                	\rho v_1v_3 & \rho v_2v_3 & \rho v_3^2 + p \\
                                                                                	\rho v_1 H & \rho v_2 H & \rho v_3 H
\end{array}\right],
\end{equation}
where $\rho$, $e$, $H$ and $p$ are the density, total energy, total enthalpy and pressure, respectively, and $\vec{v}=[v_1,v_2,v_3]^T$ is the velocity. Additionally, $\vec{F}_v$ defines the viscous fluxes,

\begin{equation}
\blocktensor{f}^{\nu}(\mu,\vec{v},\nabla\vec{v}) = \left[\begin{array}{ccc}0 & 0 & 0\\
 \tau_{xx} & \tau_{xy} & \tau_{xz} \\
 \tau_{yx} & \tau_{yy} & \tau_{yz} \\
 \tau_{zx} & \tau_{zy} & \tau_{zz} \\
 \sum_{j=1}^3 v_j\tau_{1j} + \kappa T_x& \sum_{j=1}^3 v_j\tau_{2j} + \kappa T_y& \sum_{j=1}^3 v_j\tau_{3j} + \kappa T_z
\end{array}\right],
\label{eq:viscousfluxes}
\end{equation}
where $\kappa$ is the thermal conductivity, $T_x, T_y$ and $T_z$ denote the temperature gradients and the stress tensor $\boldsymbol{\tau}$ is defined as $\boldsymbol{\tau} = \mu(\nabla \vec{v} + (\nabla \vec{v})^T) - 2/3\mu \boldsymbol{I}\nabla\cdot\vec{v}$, with $\mu$ the dynamic viscosity and $\boldsymbol{I}$ is the three-dimensional identity matrix.

\section{Error Analysis Details}
\label{ApendiceError}
In this Appendix, we develop a model for the errors arising from the described methodology. Let us define the following variables:

\begin{equation}
    \qi{NN}{} = \qm{HO}{} + \error{NN}{}
    \label{Eq:qNN}
\end{equation}
and
\begin{equation}
    \qi{T}{} = \qm{HO}{} + \error{T}{}
    \label{Eq:qT}
\end{equation}
where the subindex $NN$ denotes the solution with corrective forcing (neural network corrected), $T$ stands for the low order solution (not corrected), and the overbar means a high order filtered solution. Variables $\error{NN}{}$ and $\error{T}{}$ are the error in each case. Consider the evolution of the corrected variable:

\begin{equation}
    \frac{d \qi{NN}{}}{d t} = \bm{p}_{LO}(\qi{NN}{}) + \bm{s}(\qi{NN}{},\qi{HO}{}),
    \label{Eq:EvolPerfect}
\end{equation}
while evolution system without any correction is:

\begin{equation}
    \frac{d \qi{T}{}}{d t} = \bm{p}_{LO}(\qi{T}{}).
\end{equation}
In Eq. \eqref{Eq:EvolPerfect} and in the main text we have considered an exact corrective forcing $\bm{s}$. However, to estimate errors, we consider that the corrective forcing is not exact and has an associated error:

 \begin{equation}
     \bm{s}_{NN} = \bm{s} + \bm{e},
     \label{Eq:errorSnormal}
 \end{equation}
 We can derive an equation for the nonexact corrective forcing: 
 
\begin{equation}
    \frac{d \qi{NN}{}}{d t} = \bm{p}_{LO}(\qi{NN}{}) + \bm{s}_{NN}(\qi{NN}{}).
\end{equation}
Using both definitions of Eq. \eqref{Eq:qNN} and Eq. \eqref{Eq:qT} and introducing them into their respective evolution equations:

\begin{equation}
    \frac{d \qm{HO}{}}{d t} + \frac{d \error{T}{}}{d t}= \bm{p}_{LO}(\qm{HO}{}+\error{T}{}),
    \label{Eq:EvolErrorT1}
\end{equation}
and

\begin{equation}
    \frac{d \qm{HO}{}}{d t}+\frac{d \error{NN}{}}{d t} = \bm{p}_{LO}(\qm{HO}{}+\error{NN}{}) + \bm{s}_{NN}(\qm{HO}{}+\error{NN}{}).
    \label{Eq:EvolErrorNN1}
\end{equation}

We subtract Eq. \eqref{Eq:EvolPerfect} from Eq. \eqref{Eq:EvolErrorT1} and Eq. \eqref{Eq:EvolErrorNN1} to obtain:
\begin{equation}
    \frac{d \error{T}{}}{d t}= \bm{p}_{LO}(\qm{HO}{}+\error{T}{}) - \bm{p}_{LO}(\qm{HO}{}) - \bm{s}(\qm{HO}{}),
    \label{Eq:EvolEq1}
\end{equation}
and

\begin{equation}
    \frac{d \error{NN}{}}{d t}= \bm{p}_{LO}(\qm{HO}{}+\error{NN}{}) + \bm{s}_{NN}(\qm{HO}{}+\error{NN}{}) - \bm{p}_{LO}(\qm{HO}{}) - \bm{s}(\qm{HO}{}).
    \label{Eq:EvolEq2}
\end{equation}
In both cases, if the error is small enough (in comparison to the state variables), then it is possible to expand the equation operators $\bm{p}_{LO}$ and $\bm{s}_{NN}$ using Taylor series:

\begin{equation}
     \bm{p}_{LO}(\qm{HO}{}+\error{}{}) \approx \bm{p}_{LO}(\qm{HO}{}) + \pdv{\bm{p}_{LO}}{\qm{}{}} \error{}{} + O(\error{}{2}),
     \label{Eq:TaylorExpansionP}
\end{equation}
and

\begin{equation}
     \bm{s}_{NN}(\qm{HO}{}+\error{}{}) \approx \bm{s}_{NN}(\qm{HO}{}) + \pdv{\bm{s}_{NN}}{\qm{HO}{}} \error{}{} + O(\error{}{2}).
     \label{Eq:TaylorExpansionS}
\end{equation}
 where $\bm{e}$ is the error of the NN. 

We can now introduce Eq. \eqref{Eq:TaylorExpansionP}, Eq. \eqref{Eq:TaylorExpansionS} and Eq. \eqref{Eq:errorSnormal} into Eq. \eqref{Eq:EvolEq1} and Eq. \eqref{Eq:EvolEq2} to obtain:

\begin{equation}
    \frac{d \error{T}{}}{d t}= \pdv{\bm{p}_{LO}}{\qm{}{}} \error{T}{} - \bm{s}(\qm{HO}{}),
    \label{Eq:EvolErrorT}
\end{equation}
and

\begin{equation}
    \frac{d \error{NN}{}}{d t}= \pdv{\bm{p}_{LO}}{\qm{HO}{}} \error{NN}{} + \pdv{\bm{s}_{NN}}{\qm{HO}{}} \error{NN}{} + \bm{e}.
    \label{Eq:EvolErrorNN}
\end{equation}
Finally, we can consider the norms for Eq. \eqref{Eq:EvolErrorT} and Eq. \eqref{Eq:EvolErrorNN} to find the evolution: 

\begin{equation}
    \frac{d \norm{\error{T}{}}}{d t} \leq \norm{\pdv{\bm{p}_{LO}}{\qm{HO}{}}} \norm{\error{T}{}} + \norm{\bm{s}(\qm{HO}{})}
    \label{Eq:EvolErrorT2}
\end{equation}

\begin{equation}
    \frac{d \norm{\error{NN}{}}}{d t} \leq \norm{\pdv{\bm{p}_{LO}}{\qm{HO}{}}} \norm{\error{NN}{}}  + \norm{\pdv{\bm{s}_{NN}}{\qm{HO}{}}} \norm{\error{NN}{}} +  \norm{\bm{e}}
    \label{Eq:EvolErrorNN2}
\end{equation}

In these equations, there are two different types of terms in the equations. Those multiplied by the error itself, which are sensitivity types of error, and those that are not linearly proportional to the error. In the error from the truncated solution (Eq. \eqref{Eq:EvolErrorT2} without the NN model), we see that the error is proportional to the Jacobian of the governing equations (NS) and there is a source of error, the forcing term, which was missing in the formulation of the evolving equations. On the other hand, when modelling the forcing term with an NN in Eq. \eqref{Eq:EvolErrorNN2} we have the sensitivity of both the governing equations (NS) and also the sensitivity of the NN. In addition, there is a source of error that comes from the fact that the NN is not perfect.

We can make a simple model to predict the time evolution of the errors. Both the Jacobians from the NS equations and the NN will be time dependent in a sense that their norms will change depending on the solution and moment in time. The first simplification is to consider some ``effective" and constant coefficient $\alpha$ which for the truncated case represents the Jacobian of the NS equations but for the NN case is the Jacobian for both the NS equations and the NN. Regarding the term independent of the error, we can also introduce some ``effective" source term $\beta$, which for the NS equation is the missing source term but for the NN model is the error concurred by the NN. Eq. \eqref{Eq:EvolErrorT2} and  \eqref{Eq:EvolErrorNN2} simply to:

\begin{equation}
\frac{d \norm{\error{NN}{}}}{d t} \approx \alpha_1 \norm{\error{NN}{}} + \beta_1,
\label{Eq:ErrorSimplificado1}
\end{equation}

\begin{equation}
\frac{d \norm{\error{T}{}}}{d t} \approx \alpha_2 \norm{\error{T}{}} + \beta_2.
\label{Eq:ErrorSimplificado2}
\end{equation}
We can integrate both Eq. \eqref{Eq:ErrorSimplificado1} and Eq. \eqref{Eq:ErrorSimplificado2} to find an analytical evolution for the errors:

\begin{equation}
\norm{\error{NN}{}} \approx \frac{\beta_1}{\alpha_1} \left( e^{\alpha_1 t} - 1 \right),
\end{equation}
and

\begin{equation}
\norm{\error{T}{}} \approx \frac{\beta_2}{\alpha_2} \left( e^{\alpha_2 t} - 1 \right).
\end{equation}
We can see that both error evaluations are similar when $\alpha_1=\alpha_2$ and $\beta_1=\beta_2$. In the main text of this manuscript, we provide values for $\alpha$ and $\beta$ and explore the behaviour of these errors; see Section \ref{sec_error_main}.

\section*{Acknowledgements}
Esteban Ferrer would like to thank the support of the Spanish Minister MCIN/AEI/10.13039/501100011033 and the European Union NextGenerationEU/PRTR for the grant "Europa Investigación 2020" EIN2020-112255, and also the Comunidad de Madrid through the call Research Grants for Young Investigators from the Universidad Politécnica de Madrid. Additionally, the authors acknowledge the Universidad Politécnica de Madrid (www.upm.es) for providing computing resources on the Magerit Supercomputer.

\bibliographystyle{ieeetr} 
\bibliography{bib.bib}

\end{document}